\documentclass[notitlepage,reprint,aps,prl,floatfix,superscriptaddress,nobalancelastpage]{revtex4-1}
\usepackage{amsmath,bbm,hyperref}
\usepackage{graphicx}
\renewcommand{\vec}[1]{\mathbf{#1}}

\usepackage{color,soul}
\usepackage{siunitx}

\newcommand{\ver}{\vec{r}}

\newcommand{\ket}[1]{\lvert#1\rangle}
\newcommand{\abs}[1]{\lvert#1\rvert}
\newcommand*\dif{\mathop{}\!\mathrm{d}}
\newcommand{\e}{\mathcal{E}}

\newcommand{\citettwo}[2]{\begin{NoHyper}\citeauthor{#1}\end{NoHyper}~\cite{#1,#2}}

\begin{document}

\title{Meissner-like effect for synthetic gauge field in multimode cavity  QED}
\author{Kyle E. Ballantine}
\affiliation{SUPA, School of Physics and Astronomy, University of St. Andrews, St. Andrews KY16 9SS, United Kingdom}
\author{Benjamin L. Lev}
\affiliation{Departments of Physics and Applied Physics and Ginzton Laboratory, Stanford University, Stanford CA 94305, USA}
\author{Jonathan Keeling} 
\affiliation{SUPA, School of Physics and Astronomy, University of St. Andrews, St. Andrews KY16 9SS, United Kingdom}
\begin{abstract}
  Previous realizations of synthetic gauge fields for ultracold atoms do not allow
  the spatial profile of the field to evolve freely. We propose a scheme which
  overcomes this restriction by using the light in a multimode cavity, in
  conjunction with Raman coupling, to realize an artificial magnetic field which
  acts on a Bose-Einstein condensate of neutral atoms. We describe the
  evolution of such a system, and present the results of numerical simulations
  which show dynamical coupling between the effective field and the matter on
  which it acts. Crucially, the freedom of the spatial profile of the field is
  sufficient to realize a close analogue of the Meissner effect, where the
  magnetic field is expelled from the superfluid.  This back-action of the atoms on the synthetic field  distinguishes the Meissner-like effect described here  from the Hess-Fairbank suppression of rotation in a neutral superfluid observed elsewhere.
 \end{abstract}
\maketitle


The Meissner effect~\cite{Meissner1933} is the {\it sine qua non} of
superconductivity~\cite{LeggettBook}.  As captured by the Ginzburg-Landau
equations~\cite{Ginzburg}, the  superfluid
order parameter couples to the electromagnetic fields 
such that there is perfect diamagnetism.  Physically this
arises because the normal paramagnetic response of mater is completely
suppressed by the phase stiffness of the superfluid, leaving only the
diamagnetic current~\cite{schrieffer83}.  Magnetic field thus decays
exponentially into the bulk~\cite{London35}.  Exponentially decaying fields
are symptomatic of a massive field theory, and so can be seen as a direct
consequence of the the Anderson-Higgs mechanism~\cite{Anderson1963,Higgs1964}
giving the electromagnetic field a mass gap.  Central to all these phenomena
is that minimal coupling between the electromagnetic field and the superfluid
modifies the equations of motion for both the superfluid and the
electromagnetic field.

\begin{figure}[t!]
  \centering
   \includegraphics[width=\columnwidth]{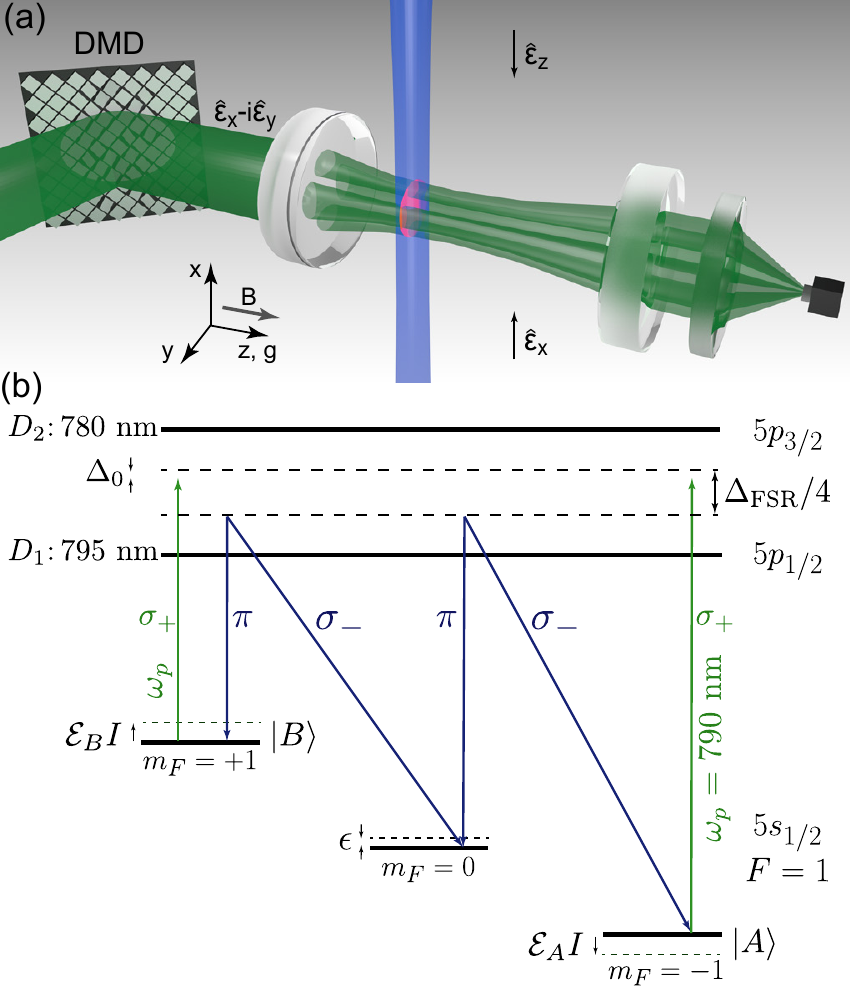}
  \caption{(a) A 2D-BEC is  coupled to linearly polarized counter-propagating Raman beams in $\hat{x}$ and trapped inside a cavity set in the multimode, confocal configuration. (Optical trapping lasers not shown, and only TEM$_{10}$ and TEM$_{01}$ modes shown for simplicity.)  The axis of the oblate BEC is collinear with the cavity axis.  The cavity, with field loss $\kappa$, is pumped along +$\hat{z}$ by a circularly polarized field of frequency $\omega_p$ detuned by $\Delta_0$ from the confocal cavity frequency $\omega_0$.  The pump laser is spatially shaped with a digital multimirror device (DMD) spatial light modulator before cavity injection.  Cavity light is imaged by an EMCCD camera. (Atomic absorption imaging laser and camera not shown.) A magnetic field $B$ is oriented along +$\hat{z}$. (b) $^{87}$Rb atomic level diagram with three coupling lasers shown.  Green arrows show the pump field that Stark shifts the state $|A\rangle$ and $|B\rangle$ by $\mathcal{E}_{A,B}$.  The laser frequency is set to the tune-out wavelength between the $D_1$ and $D_2$ lines in $^{87}$Rb. The blue arrows show the two-photon Raman-coupling fields driving  $\pi$  and $\sigma_-$ transition.  The Raman fields are detuned from  $\omega_0$ by a quarter of a free spectral range $\Delta_{FSR}/4$. }
  \label{fig1}
\end{figure}

The concept of ``synthetic'' gauge fields has attracted much attention over
the last few years.  In the context of ultracold atoms, realizations have
included schemes based on dark states~\cite{Visser1998,Juzeliunas2004} or
Raman driving~\cite{Spielman2009,Lin2009a,Lin2009b}, or inducing Peierls
phases in lattice systems~\cite{Jaksch2003,Sorensen2005,
  Kolovsky2011,Cooper2011,Struck2011,Aidelsburger2011,Hauke2012,Struck2012,
  Miyake2013,Aidelsburger2013,atala14} (for a review,
see~\cite{Dalibard2011,Goldman2014,Zhai12,Zhai15}).  There have also been proposals to
realize gauge fields for photons, including ``free space'' realizations using
Rydberg atoms in non-planar ring cavity geometries~\cite{Schine2016} as well
as Peierls phases for photon hopping in coupled cavity
arrays~\cite{Fang2012,Hafezi2013,Dubcek2015}.  However, with a few exceptions,
all these have involved static gauge fields---there is no feedback of the
atoms (or photons) on the synthetic field.  Thus, even in the pioneering
demonstration of a Meissner phase of chiral currents~\cite{atala14}, it is
noted that these experiments are closer to the Hess-Fairbank effect~\cite{Hess67}
(suppression of rotation in a neutral superfluid), and do not show
expulsion of the synthetic field.  In contrast, a charged superfluid acts back on the magnetic field.

The  synthetic field cannot be expelled in the above schemes because it is
set by a fixed external laser or the system geometry.  The
exceptions  are thus proposals where the strength of synthetic field
depends on a dynamical quantity.  One such proposal is in optomechanical
cavity arrays with a Peierls phase for photon hopping set
by mechanical oscillators~\cite{Walter2015}.  Another proposal is to consider atoms in optical
cavities,  replacing the external laser drive by light in the cavity,
as has been recently proposed by~\citet{Zheng2016}, with a two photon-assisted
hopping scheme involving the cavity and a transverse pump.  This scheme 
naturally relates to the self-organization of atoms under transverse
pumping~\cite{Domokos2002,Black2003,Baumann2010}, and proposed extensions
involving spin-orbit 
coupling~\cite{deng14,dong14,padhi14,pan15,Dong2015} and self-organized chiral
states~\cite{Kollath2016,Sheikhan2016}.  However, in these schemes, a single-mode cavity is used, giving a ``mean-field'' coupling due to the 
infinite-range nature of the interactions, and not the local  coupling to the
gauge potential present in the Ginzburg-Landau equations.

Locality can be restored in a multimode cavity---in a cavity that
supports multiple nearly degenerate modes one can build localized
wavepackets~\cite{Gopalakrishnan2009}.  This  is also  illustrated for a longitudinally pumped system when considering the Talbot
effect~\cite{talbot36} with cold atoms~\cite{Ackemann2001}:  When atoms are 
placed in front of a planar mirror and illuminated with coherent light,
phase modulation of the light is transformed by
propagation into intensity modulation, leading to self-organization.  This
effect can be viewed as an effective atom-atom interaction mediated by
light, leading to density-wave
formation~\cite{Labeyrie2014,Diver2014,Robb2015}.   In this Letter, we show how
multimode cavity QED~\cite{Gopalakrishnan2009,Gopalakrishnan2010,Wickenbrock2013,Kollar2015,Kollar2016} can be used to realize a dynamical gauge field for cold
atoms capable of realizing an analogue of the Meissner effect for
charged superfluids.  The
simultaneous presence of near-degenerate cavity modes allows the spatial
intensity profile of the cavity light field to change over time in response to
the state of the atoms, in a form directly analogous to the Ginzburg-Landau
equations.

By realizing such a multimode cavity QED simulator of matter and dynamical gauge fields,
numerous possibilities arise.  Most intriguingly, the tunability of the
parameters controlling the effective matter-light coupling potentially allow
one to simulate the behaviour of matter with alternate values of the fine
structure constant $\alpha$.  This could provide a continuum gauge field
theory simulator complementary to the proposals for simulating lattice gauge
field theories using ultracold
atoms~\cite{Tewari2006,Cirac2010,Zohar2011,Zohar2012,Banerjee2012,
  Banerjee2013,Edmonds2013,Zohar2013a,Zohar2013b,Kasamatsu2013,Zohar2016},
trapped ions~\cite{Hauke2013,Tagliacozzo2013,Yang2016,martinez16}, or
superconducting circuits~\cite{Marcos2013,Marcos2014}.  Other opportunities
arising from our work are to explore the differences between Bose-condensed,
thermal, and fermionic atoms in the geometry considered above: The Meissner
effect depends on the phase stiffness of superfluid atoms, so should vanish at
higher temperatures.  For fermions, one might use synthetic field expulsion as
a probe of BCS superfluidity.
 
Our proposal is based on the
confocal cavity system recently realized by~\citettwo{Kollar2015}{Kollar2016}.
This cavity may be tuned between confocal multimode and single-mode
configurations.  In addition, light can be pumped transversely or
longitudinally, and patterned using a digital light
modulator~\cite{Papageorge2016}.
Multimode cavity QED has previously been proposed to explore
beyond mean-field physics in the self-organization of ultracold
atoms~\cite{Gopalakrishnan2009,Gopalakrishnan2010,Kollar2016}.

%

%
We consider this cavity in the near-confocal case, so that many
modes are near-degenerate. This cavity contains a 2D-condensate
of atoms confined along the cavity axis as shown in Fig.~\ref{fig1}. 
These atoms have two low-lying internal states, $\ket{A}$ and $\ket{B}$, 
which are coupled by two counter-propagating Raman beams via a higher
intermediate state. For example, these could be spin states 
of $^{87}\mathrm{Rb}$ split by a magnetic field~\cite{Spielman2009,Lin2009b}. 
If the population of the excited state is negligible, the Raman beams lead to
an effective coupling $\Omega$ between the two lower states. An atom in state 
$\ket{A}$ gains momentum $q\hat{x}$ by absorbing a photon from one transverse 
beam,  and loses momentum $-q\hat{x}$ by emitting a photon into the other 
beam, finishing in state $\ket{B}$, where
$q=2\pi/\lambda$ is the momentum of the Raman beams. Hence,
the two states have momentum differing by $2q\hat{x}$. Crucially, each
state has a different Stark shift due to the cavity
intensity $I$, with coefficients $\e_{A,B}$.   This gives
an atomic Hamiltonian of the form:
$  \hat{H}_\text{atom} = 
  \int d^2 \vec{r} \dif z
  \tilde{\Psi}^\dagger_{\vec{r},z}
  [\vec{h}
  +
  V_\text{ext}(\vec r,z)  ]
  \tilde{\Psi}^{}_{\vec{r},z}
  +
   \hat{H}_{\text{int}},
$
where
  \begin{equation}
  \vec{h} =\!
  \begin{pmatrix}
    \frac{(\!-i\nabla-q\hat{\vec{x}})^2}{2m}  -\e_{\!A} I(\ver,\!z) & \Omega/2 \\ 
    \Omega/2 &  \frac{(\!-i\nabla+q\hat{\vec{x}})^2}{2m} - \e_{\!B} I(\ver,\!z)
  \end{pmatrix}\!.
  \end{equation}
For simplicity, here and in the following,
we write $\vec{r}=(x,y)$, write the $z$ dependence separately,
and set $\hbar=1$.  The term $\hat{H}_{\text{int}}$ describes contact
interactions between atoms with strength $U$.

The cavity pump field wavelength is set to be at the ``tune-out" point, 780.018~nm, between the $D_1$ and $D_2$ lines in $^{87}$Rb~\cite{Schmidt:2016hv}.  The scalar light shift is zero at this wavelength, which means that the atom trapping frequencies are nearly unaffected by the cavity light in the absence of Raman coupling. The vector light shift is approximately equal and opposite for states $|A\rangle$ and $|B\rangle$ due to their opposite $m_F$ projections, i.e., $\e_A=-\e_B$.  Spontaneous emission is low, less than $\sim$10 Hz for the required Stark shifts, since this light is far detuned from either excited state.  The Raman coupling scheme is identical to those in Refs.~\cite{Lin2009b,LeBlanc2015} that exhibited nearly 0.5-s spontaneous-emission-limited BEC lifetimes. The Raman lasers do not scatter into the cavity, since they are detuned $\Delta_\text{FSR}/4 = 3.25$~GHz from any family of degenerate cavity modes for a $L = 1$-cm confocal cavity~\cite{Kollar2015}.  The BEC lifetime under our cavity and Raman-field dressing scheme should therefore be more than 100~ms, sufficient to observe the predicted physics.

The artificial magnetic field that arises from the Raman driving
scheme is in $\hat{z}$~\cite{Spielman2009}, and so the
interesting atomic dynamics will be in the transverse ($\hat{x}$-$\hat{y}$)
plane.  We therefore consider a 2D pancake of atoms, with strong trapping in
$\hat{z}$, such that we may write $\Psi_{A,B}(\ver,z)=\psi_{A,B}(\ver) Z(z)$,
where $Z(z)$ is a narrow Gaussian profile due to the strong trapping in
$\hat{z}$. In this strong trapping limit, we can integrate out the $z$
dependence to produce effective equations of motion for the transverse
wavefunctions $\psi_{A,B}$ and the transverse part of the cavity light field
$\varphi$. We consider the case where the atom cloud is trapped near one end of
the cavity, $z_0 \simeq \pm z_R$, as this leads to a quasi-local coupling
between atoms and cavity light (see
supplemental material~\footnote[1]{Supplemental material, containing details of reduction to effective 2D equations, details of numerical simulation, and discussion of artificial magnetic field strength suppression in the geometry under consideration.} for details). We choose to normalize the atomic
wavefunctions such that $\int \dif^2 \ver\,
\left(\abs{\psi_A}^2+\abs{\psi_B}^2\right)=1$ so that the number of atoms $N$
appears explicitly.  The transverse equations of motion then take the form:
\begin{widetext}
\begin{align}
  \label{phieom}
i\partial_t \varphi &= \left[
\frac{\delta}{2}\left(-l^2 \nabla^2 + \frac{r^2}{l^2}\right) -\Delta_0 - i\kappa -N\e_{\Sigma}\left(\abs{\psi_A}^2+\abs{\psi_B}^2\right) 
-N\e_{\Delta}(\abs{\psi_A}^2-\abs{\psi_B}^2)\right]\varphi  + f(\vec{r}),\\
i\partial_t
\begin{pmatrix} \psi_A \\ \psi_B \end{pmatrix} 
&=\left[-\frac{\nabla^2}{2m}+V_\text{ext}(\vec{r})
-\e_{\Sigma} \abs{\varphi}^2 + NU \left(\abs{\psi_A}^2+\abs{\psi_B}^2\right)
+ \begin{pmatrix}
    -\e_\Delta \abs{\varphi}^2  + i \frac{q}{m} \partial_x  & \Omega/2 \\ 
    \Omega/2 & \e_\Delta \abs{\varphi}^2 - i \frac{q}{m} \partial_x
  \end{pmatrix}\right]
  \begin{pmatrix} \psi_A \\ \psi_B \end{pmatrix}.
  \label{psieom}
\end{align}
\end{widetext}
Here we have rewritten $\e_\Sigma=(\e_A+\e_B)/4$ and
$\e_\Delta=(\e_A-\e_B)/4$.  $\Delta_0=\omega_P-\omega_0$ is the detuning of
the pump from the confocal cavity frequency $\omega_0$, and
$f(\ver)$ is the pump profile. We are considering modes in a nearly confocal cavity
for the cavity field $\varphi$~\cite{Kollar2015}. 
In such a case, a given family
of nearly degenerate modes takes the form of either even or odd Gauss-Hermite
functions, where $\sqrt{2}l$ is the beam waist~\cite{Siegman1986a}.  
The first term in Eq.~(\ref{phieom}) is the real-space operator describing 
the splitting between the nearly degenerate modes when the cavity is detuned
away from confocality with mode splitting $\delta$.
The restriction to only even modes means
we must restrict to $\varphi(\vec r) = \varphi(-\vec r)$; however, 
Eqs.~(\ref{phieom}) and~(\ref{psieom}) preserve this symmetry if it is initially
present.

We first describe in general terms the behavior we can expect from
Eqs.~(\ref{phieom}) and~(\ref{psieom}) before discussing the full steady-state
of these equations. As this model is closely related to that proposed by
\citet{Spielman2009}, one may expect the same behavior to occur~\cite{Lin2009b}. 
Namely, if we construct a basis transformation between the original levels and
the dressed states $(\psi_+,\psi_-)^T = U(\psi_A,\psi_B)^T$ such that the
atomic Hamiltonian is diagonalized, then atoms in each of these dressed states
see effective (opposite) magnetic fields.
Ignoring atom-atom interactions, the
dispersion relation for the lower manifold is
\begin{equation}\label{Happrox}
E_-(\vec{k}) = \frac{1}{2m^\ast}\left(\vec{k}-Q \abs{\varphi}^2\hat{\mathbf{x}}\right)^2
-\e_\Sigma\abs{\varphi}^2-\e_{4}\abs{\varphi}^4.
\end{equation}
This can be recognized as the energy of 
a particle in a magnetic vector potential
$\vec{A}=\abs{\varphi}^2 \hat{\vec{x}}$ with effective mass 
$m^\ast=m\Omega/(\Omega-2q^2/m)$ and charge 
$Q=2\e_\Delta q/(\Omega-2q^2/m)$,
as well as additional geometric scalar potential terms characterized by 
$\e_\Sigma$ and $\e_{4}=\e_\Delta^2/(\Omega-2q^2/m)$. 

As compared to Refs.~\cite{Spielman2009,Lin2009b}, the crucial difference in
Eqs.~(\ref{phieom}) and~(\ref{psieom}) is that the atomic dynamics also acts back on
the cavity light field $\varphi$.  We can explore the nature of this
back-action by expanding the low-energy eigenstate $\psi_-$ to first order in
$1/\Omega$. In the low-energy manifold, 
the difference $\abs{\psi_A}^2-\abs{\psi_B}^2$ in
Eq.~(\ref{phieom}) can be related to the wavefunction $\psi_-$ by
\begin{multline}
\label{feedback}
\abs{\psi_A}^2-\abs{\psi_B}^2 = 
\frac{q}{im\Omega}\left(\psi_{-}^\ast\partial_x\psi_{-} - \psi_{-}\partial_x\psi_{-}^\ast\right) \\ +\frac{2\e_\Delta}{\Omega}\abs{\psi_{-}}^2\abs{\varphi}^2.
\end{multline}
This  has exactly the expected form of the effect of charged
particles on a vector potential: there is a current dependent term,
the first term, followed by a diagmagnetic term.  
The current dependent term appearing in 
Eq.~(\ref{phieom}) would lead to a paramagnetic response.
However in a superfluid state, this contribution is suppressed due to the
phase stiffness of the superfluid,
 and the surviving diamagnetic term then leads to the Meissner effect.  

\begin{figure}[!htbp]
  \centering
   \includegraphics[width=\columnwidth]{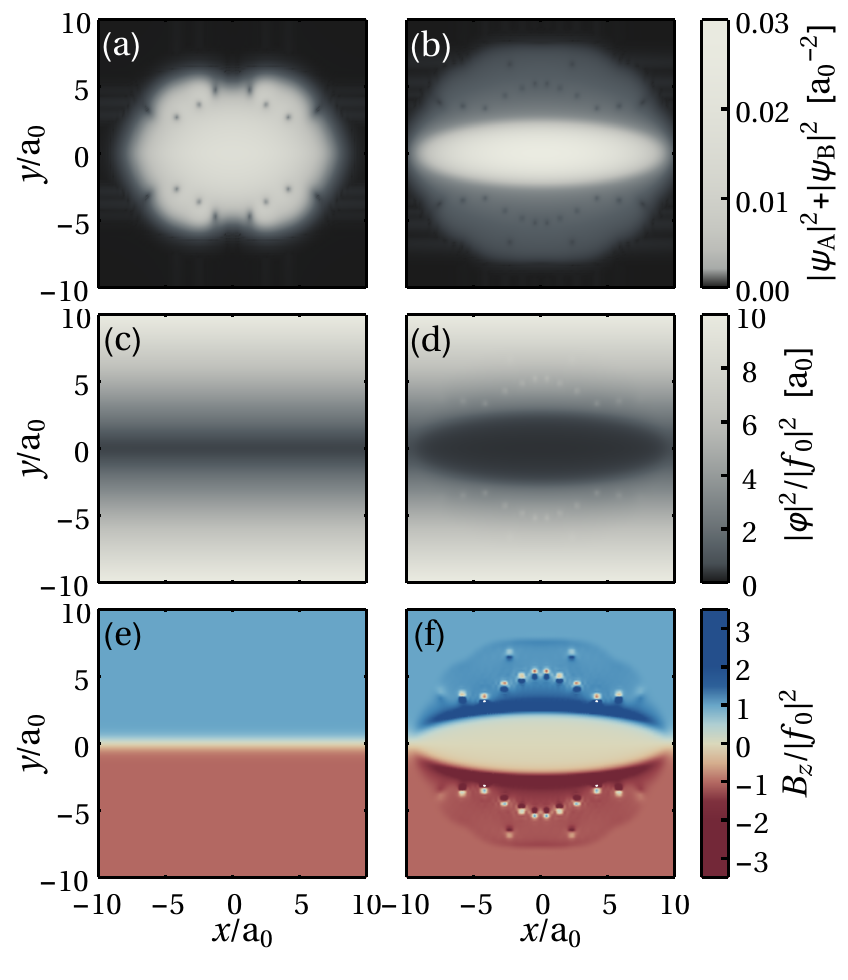}
  \caption{(a) Ground state of condensate in a static field $\abs{\varphi}^2\propto\abs{y}$ showing vortex formation. (b) Density of condensate when the field is allowed to evolve.
  No vortices remain in the cloud.
    (c) Relative intensity (i.e., $\abs{\varphi}^2/\abs{f_0}^2$) of light in cavity with no atom-light coupling used to generate (a). (d) Steady-state  relative intensity of light when coupled to atomic cloud, showing reduction in region where atoms are present. Vortices remain in region of low atomic density where the intensity recovers and so the magnetic field is high. (e) Applied magnetic field (derivative of intensity in (c)) and
    (f) magnetic field after coupled evolution showing the applied field has been
    expelled from the condensate.
  All lengths are in units of  the harmonic oscillator length $a_0=1/\sqrt{m\omega_x}$.  Using $\omega_x$ as the frequency unit, other parameters are given 
  by $\delta=-10$, $l=1$, $\Delta_0=-50$, $\kappa=1000$, $\e_\Delta=50$, 
  $q=70$,  $\Omega=10^5$, $mU=1.5\times10^{-3}$, $N=10^6$, and $f_0=4$, all relevant for cavity considered in~\cite{Kollar2016,Kollar2015}. }
  \label{fig2}
\end{figure}

\begin{figure}[!htbp]
  \centering
   \includegraphics[width=\columnwidth]{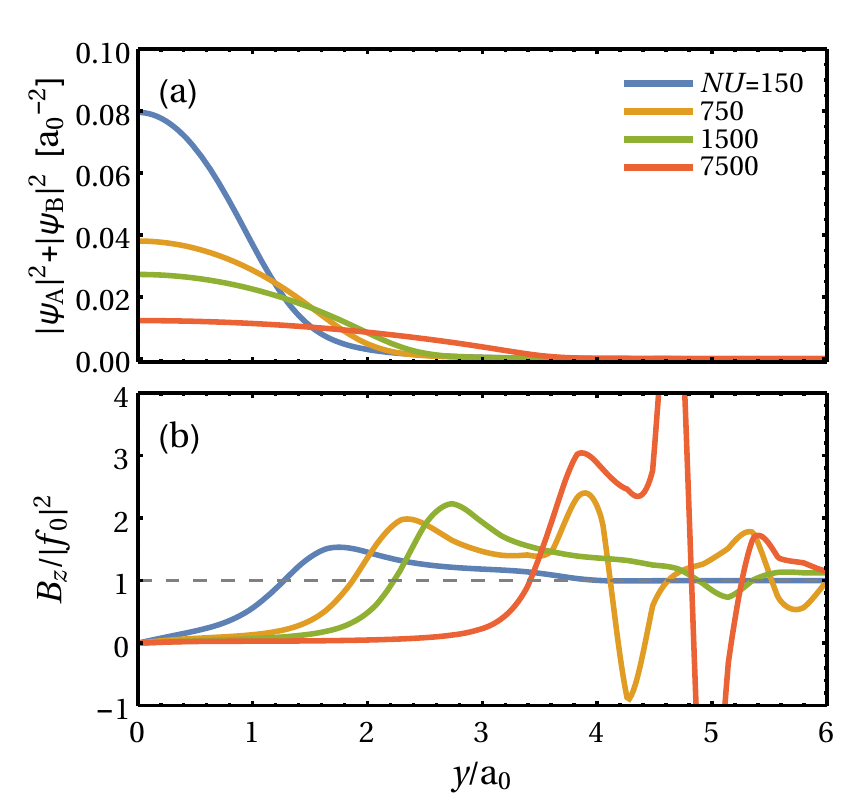}
  \caption{ Profile of (a) atomic density and (b) artificial field with varying number of atoms, in terms of the dimensionless parameter $mNU$. For realistic values, this corresponds to $N=10^5$ (blue), $N=5\times 10^5$ (yellow), $N=10^6$ (green), $N=5\times 10^6$ (orange). All other parameters are the same as in Fig.~\ref{fig2}. }
  \label{fig3}
\end{figure}

This diamagnetic response is shown in Fig.~\ref{fig2}.  Here we consider a
source term $f(\vec{r})$, the pump field, which, in the absence of atoms, we set to have
an intensity profile such that $|\varphi|^2 \propto |y|$, 
and thus the magnetic field has uniform magnitude, but with a sign dependent
on $y$ (see supplemental material~\footnotemark[1]). 
The applied effective magnetic field is 
$B_z=\partial_y\abs{\varphi}^2\approx\pm |f_0|^2$, where $f_0$ is a constant
which specifies the overall amplitude of the pump.
The rows of Fig.~\ref{fig2} show the atomic density,
the cavity intensity and the artificial magnetic field, respectively. The
left column shows the case where the field is artificially kept static, i.e., 
where the terms proportional
to $\psi_{A,B}$ are omitted from Eq.~(\ref{phieom}) to prevent any feedback, and
the right column  shows the case when the field is allowed to evolve.  

As expected, in Fig.~\ref{fig2}(f) we see that with the feedback included,
the magnetic field is suppressed in the region containing the atomic
cloud. Compared to the applied field (e), the field is suppressed in
a large area  where the atomic density, shown in (a), is high.
This is a consequence of the change in cavity field which goes from being linear in $|y|$
as shown in (c), to having a region of near constant value shown in (d).  
Additionally, as the cavity light intensity
recovers to its default value, this causes an increase in its
gradient, and so in panel (f) one sees the artificial magnetic field is
increased immediately outside the atom cloud.
In response to the changing field, we see that while
in panel  (a) there are vortices due to the static field,
in panel (b) the higher density region no longer
contains vortices, although they remain around the edges of the where
there is still high magnetic field.
This reduction in field $|\varphi|$ also leads to a change in the geometric
scalar potential terms in
Eq.~(\ref{Happrox}), which is what causes the condensate to shrink in size (see
also below). 
Further out, the condensate density decays slowly in $\hat{y}$,
and the remaining vortices lead to small modulations in the intensity.
However, these modulations over small areas lead to quite large
derivatives which show up as flux-antiflux pairs. 

While the results in Fig.~\ref{fig2} clearly show suppression of the magnetic field analogous to the Meissner effect, it is important to note several differences between Eqs.~(\ref{phieom}) and~(\ref{psieom}) and the standard Meissner effect.  In particular, while the atoms see an effective vector potential $|\varphi|^2 \hat{\vec{x}}$, the remaining action for the field $|\varphi|^2$ does not simulate the Maxwell action for a gauge field: the remaining action does not have gauge symmetry, and has a small residual gap (due to the cavity loss and detuning).  Because the action for $\varphi$ is not  gapless, there is already exponential decay of the synthetic field away from its source.  What our results show is that coupling to a superfluid significantly enhances the gap for this field (in our case, by at least one order of magnitude), thus leading to a significant suppression of the field in the region where the atoms live.  These points are discussed further in~\footnotemark[1].

The figures are the result of numerical simulations of 
Eqs.~(\ref{phieom}) and~(\ref{psieom}) performed by
using XMDS2 (eXtensible Multi-Dimensional Simulator)~\cite{Dennis2013201}. 
They represent the steady-state which is reached
when the artificial magnetic field, proportional to the pump amplitude, 
is adiabatically increased from zero. 
A computationally efficient way to find this is to evolve the equation of
motion for the light field, which includes pumping and loss, in real time,
while simultaneously evolving the equation of motion for the condensate
in imaginary time and renormalizing this wavefunction at each step.
Hence, we find a state which is both a steady-state of the
real-time equations of motion and also the ground state for the atoms in
that given intensity profile. This method correctly finds the steady-state
profile, but the transient dynamics does not match that which would
be seen experimentally;  as we focus only here on the steady-states, this
is not a problem but will be addressed in future work.
We make use of the natural
units of length and frequency set by the hamonic trap, i.e., 
$a_0=1/\sqrt{m\omega_x}$ and $\omega_x$. For realistic experimental parameters,
$\omega_x$ may be on the order of $\SI{1}{\kilo\hertz}$ and 
$a_0\approx 1$~$\mu$m.
Values of all other parameters are given in the figure captions.

The change in shape of the condensate in Fig.~\ref{fig2}(c) can be explained
by our choice of potential $V_\mathrm{ext}(\ver)$. 
As is clear from Eq.~(\ref{Happrox}), the field $\varphi$ leads both
to vector and  scalar potential terms.  The scalar potentials
have two contributions: quadratic and quartic.
As a consequence of working at the tune-out wavelength, the quadratic scalar potentials in Eq.~(\ref{Happrox}) are zero, i.e.,  $\e_\Sigma=0$.  The quartic term is
harder to eliminate, but its effects can be mitigated.
Part of the effect of the quartic term is to lead to a reduction (or even reversal) 
of the transverse harmonic trapping of the atoms.  This can
be compensated by choosing a trap of the form
\begin{equation} 
V_\text{ext} = \frac{m\omega_x^2}{2}\left(x^2+y^2\right)+\e_4 \abs{f_0}^4 y^2,
\end{equation}
i.e., an asymmetric harmonic oscillator potential
with $\omega_y=\sqrt{\omega_x^2+2\e_4\abs{f_0}^2/m}$. However, this expression
assumes $|\varphi|^2=\abs{f_0}^2 |y|$, as would hold
in the absence of atomic diamagnetism. 
As the magnetic field is expelled, the scalar potential also changes, and
the condensate becomes more tightly trapped.

Figure~\ref{fig3} shows cross-sections at $x=0$ of (a) the atomic density
and (b) the artificial magnetic field for various values of the total
number of atoms. Increasing the number of atoms leads to a larger
condensate with a lower peak density (integral of density remains normalized). 
The resulting magnetic field profile is shown in (b). The area of low
magnetic field is closely aligned to the area of high atomic density. At the
edges of the cloud, the artificial field is increased
as the expelled field accumulates here, before returning to its default value
well outside the cloud. The effect of vortices in the very low density areas
can also be seen.

In summary, we have shown how a spatially varying synthetic gauge field can be achieved, and how magnetc field suppression in an atomic superfluid, analogous to the Meissner effect, can be realized.  By effecting an artifical magnetic field proportional to  the intensity of light in a multimode cavity, we have coupled the dynamics of the spatial profile of the magnetic field and the atomic wavefunction. Our results illustrate the potential of multimode cavity QED to simulate dynamical gauge fields, and to explore self-consistent steady states, and potentially phase transitions, of matter coupled to synthetic gauge fields.

\begin{acknowledgments}
  We acknowledge helpful discussions with
  E.~Altman,  M.~J.~Bhaseen, S.~Gopalakrishnan, B.~D.~Simons, and I.~Spielman.  We thank V.~D.~Vaidya and Y.~Guo for help in preparation of the manuscript.
  K.E.B.\ and J.K.\ acknowledge support from EPSRC program “TOPNES”
  (EP/I031014/1). J.K.\ acknowledges support from the Leverhulme Trust
  (IAF-2014-025). B.L.L.\ acknowledges support from the ARO and the David
  and Lucille Packard Foundation.
\end{acknowledgments}


\begin{thebibliography}{77}%
\makeatletter
\providecommand \@ifxundefined [1]{%
 \@ifx{#1\undefined}
}%
\providecommand \@ifnum [1]{%
 \ifnum #1\expandafter \@firstoftwo
 \else \expandafter \@secondoftwo
 \fi
}%
\providecommand \@ifx [1]{%
 \ifx #1\expandafter \@firstoftwo
 \else \expandafter \@secondoftwo
 \fi
}%
\providecommand \natexlab [1]{#1}%
\providecommand \enquote  [1]{``#1''}%
\providecommand \bibnamefont  [1]{#1}%
\providecommand \bibfnamefont [1]{#1}%
\providecommand \citenamefont [1]{#1}%
\providecommand \href@noop [0]{\@secondoftwo}%
\providecommand \href [0]{\begingroup \@sanitize@url \@href}%
\providecommand \@href[1]{\@@startlink{#1}\@@href}%
\providecommand \@@href[1]{\endgroup#1\@@endlink}%
\providecommand \@sanitize@url [0]{\catcode `\\12\catcode `\$12\catcode
  `\&12\catcode `\#12\catcode `\^12\catcode `\_12\catcode `\%12\relax}%
\providecommand \@@startlink[1]{}%
\providecommand \@@endlink[0]{}%
\providecommand \url  [0]{\begingroup\@sanitize@url \@url }%
\providecommand \@url [1]{\endgroup\@href {#1}{\urlprefix }}%
\providecommand \urlprefix  [0]{URL }%
\providecommand \Eprint [0]{\href }%
\providecommand \doibase [0]{http://dx.doi.org/}%
\providecommand \selectlanguage [0]{\@gobble}%
\providecommand \bibinfo  [0]{\@secondoftwo}%
\providecommand \bibfield  [0]{\@secondoftwo}%
\providecommand \translation [1]{[#1]}%
\providecommand \BibitemOpen [0]{}%
\providecommand \bibitemStop [0]{}%
\providecommand \bibitemNoStop [0]{.\EOS\space}%
\providecommand \EOS [0]{\spacefactor3000\relax}%
\providecommand \BibitemShut  [1]{\csname bibitem#1\endcsname}%
\let\auto@bib@innerbib\@empty
\bibitem [{\citenamefont {Meissner}\ and\ \citenamefont
  {Ochsenfeld}(1933)}]{Meissner1933}%
  \BibitemOpen
  \bibfield  {author} {\bibinfo {author} {\bibfnamefont {W.}~\bibnamefont
  {Meissner}}\ and\ \bibinfo {author} {\bibfnamefont {R.}~\bibnamefont
  {Ochsenfeld}},\ }\href {\doibase 10.1007/BF01504252} {\bibfield  {journal}
  {\bibinfo  {journal} {Naturwissenschaften}\ }\textbf {\bibinfo {volume}
  {21}},\ \bibinfo {pages} {787} (\bibinfo {year} {1933})}\BibitemShut
  {NoStop}%
\bibitem [{\citenamefont {Leggett}(2006)}]{LeggettBook}%
  \BibitemOpen
  \bibfield  {author} {\bibinfo {author} {\bibfnamefont {A.}~\bibnamefont
  {Leggett}},\ }\href {https://books.google.co.uk/books?id=PiRRAAAAMAAJ} {\emph
  {\bibinfo {title} {Quantum Liquids: Bose Condensation and Cooper Pairing in
  Condensed-matter Systems}}},\ Oxford graduate texts in mathematics\ (\bibinfo
   {publisher} {OUP Oxford},\ \bibinfo {year} {2006})\BibitemShut {NoStop}%
\bibitem [{\citenamefont {Ginzburg}\ and\ \citenamefont
  {Landau}(1950)}]{Ginzburg}%
  \BibitemOpen
  \bibfield  {author} {\bibinfo {author} {\bibfnamefont {V.~L.}\ \bibnamefont
  {Ginzburg}}\ and\ \bibinfo {author} {\bibfnamefont {L.~D.}\ \bibnamefont
  {Landau}},\ }\href@noop {} {\bibfield  {journal} {\bibinfo  {journal} {Zh.
  Eksp. Teor. Fiz}\ }\textbf {\bibinfo {volume} {20}},\ \bibinfo {pages} {1064}
  (\bibinfo {year} {1950})}\BibitemShut {NoStop}%
\bibitem [{\citenamefont {Schrieffer}(1983)}]{schrieffer83}%
  \BibitemOpen
  \bibfield  {author} {\bibinfo {author} {\bibfnamefont {J.~R.}\ \bibnamefont
  {Schrieffer}},\ }\href@noop {} {\emph {\bibinfo {title} {Theory of
  Superconductivity}}}\ (\bibinfo  {publisher} {Westview Press},\ \bibinfo
  {address} {Boulder, CO},\ \bibinfo {year} {1983})\BibitemShut {NoStop}%
\bibitem [{\citenamefont {London}\ and\ \citenamefont
  {London}(1935)}]{London35}%
  \BibitemOpen
  \bibfield  {author} {\bibinfo {author} {\bibfnamefont {F.}~\bibnamefont
  {London}}\ and\ \bibinfo {author} {\bibfnamefont {H.}~\bibnamefont
  {London}},\ }\href {\doibase 10.1098/rspa.1935.0048} {\bibfield  {journal}
  {\bibinfo  {journal} {Proc. R. Soc. A}\ }\textbf {\bibinfo {volume} {149}},\
  \bibinfo {pages} {71} (\bibinfo {year} {1935})}\BibitemShut {NoStop}%
\bibitem [{\citenamefont {Anderson}(1963)}]{Anderson1963}%
  \BibitemOpen
  \bibfield  {author} {\bibinfo {author} {\bibfnamefont {P.~W.}\ \bibnamefont
  {Anderson}},\ }\href {\doibase 10.1103/PhysRev.130.439} {\bibfield  {journal}
  {\bibinfo  {journal} {Phys. Rev.}\ }\textbf {\bibinfo {volume} {130}},\
  \bibinfo {pages} {439} (\bibinfo {year} {1963})}\BibitemShut {NoStop}%
\bibitem [{\citenamefont {Higgs}(1964)}]{Higgs1964}%
  \BibitemOpen
  \bibfield  {author} {\bibinfo {author} {\bibfnamefont {P.~W.}\ \bibnamefont
  {Higgs}},\ }\href {\doibase 10.1103/PhysRevLett.13.508} {\bibfield  {journal}
  {\bibinfo  {journal} {Phys. Rev. Lett.}\ }\textbf {\bibinfo {volume} {13}},\
  \bibinfo {pages} {508} (\bibinfo {year} {1964})}\BibitemShut {NoStop}%
\bibitem [{\citenamefont {Visser}\ and\ \citenamefont
  {Nienhuis}(1998)}]{Visser1998}%
  \BibitemOpen
  \bibfield  {author} {\bibinfo {author} {\bibfnamefont {P.~M.}\ \bibnamefont
  {Visser}}\ and\ \bibinfo {author} {\bibfnamefont {G.}~\bibnamefont
  {Nienhuis}},\ }\href {\doibase 10.1103/PhysRevA.57.4581} {\bibfield
  {journal} {\bibinfo  {journal} {Phys. Rev. A}\ }\textbf {\bibinfo {volume}
  {57}},\ \bibinfo {pages} {4581} (\bibinfo {year} {1998})}\BibitemShut
  {NoStop}%
\bibitem [{\citenamefont {Juzeli\ifmmode~\bar{u}\else \={u}\fi{}nas}\ and\
  \citenamefont {\"Ohberg}(2004)}]{Juzeliunas2004}%
  \BibitemOpen
  \bibfield  {author} {\bibinfo {author} {\bibfnamefont {G.}~\bibnamefont
  {Juzeli\ifmmode~\bar{u}\else \={u}\fi{}nas}}\ and\ \bibinfo {author}
  {\bibfnamefont {P.}~\bibnamefont {\"Ohberg}},\ }\href {\doibase
  10.1103/PhysRevLett.93.033602} {\bibfield  {journal} {\bibinfo  {journal}
  {Phys. Rev. Lett.}\ }\textbf {\bibinfo {volume} {93}},\ \bibinfo {pages}
  {033602} (\bibinfo {year} {2004})}\BibitemShut {NoStop}%
\bibitem [{\citenamefont {Spielman}(2009)}]{Spielman2009}%
  \BibitemOpen
  \bibfield  {author} {\bibinfo {author} {\bibfnamefont {I.~B.}\ \bibnamefont
  {Spielman}},\ }\href@noop {} {\bibfield  {journal} {\bibinfo  {journal}
  {Phys. Rev. A}\ }\textbf {\bibinfo {volume} {79}},\ \bibinfo {pages} {063613}
  (\bibinfo {year} {2009})}\BibitemShut {NoStop}%
\bibitem [{\citenamefont {Lin}\ \emph {et~al.}(2009{\natexlab{a}})\citenamefont
  {Lin}, \citenamefont {Compton}, \citenamefont {Perry}, \citenamefont
  {Phillips}, \citenamefont {Porto},\ and\ \citenamefont
  {Spielman}}]{Lin2009a}%
  \BibitemOpen
  \bibfield  {author} {\bibinfo {author} {\bibfnamefont {Y.-J.}\ \bibnamefont
  {Lin}}, \bibinfo {author} {\bibfnamefont {R.~L.}\ \bibnamefont {Compton}},
  \bibinfo {author} {\bibfnamefont {A.~R.}\ \bibnamefont {Perry}}, \bibinfo
  {author} {\bibfnamefont {W.~D.}\ \bibnamefont {Phillips}}, \bibinfo {author}
  {\bibfnamefont {J.~V.}\ \bibnamefont {Porto}}, \ and\ \bibinfo {author}
  {\bibfnamefont {I.~B.}\ \bibnamefont {Spielman}},\ }\href {\doibase
  10.1103/PhysRevLett.102.130401} {\bibfield  {journal} {\bibinfo  {journal}
  {Phys. Rev. Lett.}\ }\textbf {\bibinfo {volume} {102}},\ \bibinfo {pages}
  {130401} (\bibinfo {year} {2009}{\natexlab{a}})}\BibitemShut {NoStop}%
\bibitem [{\citenamefont {Lin}\ \emph {et~al.}(2009{\natexlab{b}})\citenamefont
  {Lin}, \citenamefont {Compton}, \citenamefont {Jimenez-Garcia}, \citenamefont
  {Porto},\ and\ \citenamefont {Spielman}}]{Lin2009b}%
  \BibitemOpen
  \bibfield  {author} {\bibinfo {author} {\bibfnamefont {Y.-J.}\ \bibnamefont
  {Lin}}, \bibinfo {author} {\bibfnamefont {R.~L.}\ \bibnamefont {Compton}},
  \bibinfo {author} {\bibfnamefont {K.}~\bibnamefont {Jimenez-Garcia}},
  \bibinfo {author} {\bibfnamefont {J.~V.}\ \bibnamefont {Porto}}, \ and\
  \bibinfo {author} {\bibfnamefont {I.~B.}\ \bibnamefont {Spielman}},\ }\href
  {\doibase 10.1038/nature08609} {\bibfield  {journal} {\bibinfo  {journal}
  {Nature (London)}\ }\textbf {\bibinfo {volume} {462}},\ \bibinfo {pages}
  {628} (\bibinfo {year} {2009}{\natexlab{b}})}\BibitemShut {NoStop}%
\bibitem [{\citenamefont {Jaksch}\ and\ \citenamefont
  {Zoller}(2003)}]{Jaksch2003}%
  \BibitemOpen
  \bibfield  {author} {\bibinfo {author} {\bibfnamefont {D.}~\bibnamefont
  {Jaksch}}\ and\ \bibinfo {author} {\bibfnamefont {P.}~\bibnamefont
  {Zoller}},\ }\href {http://stacks.iop.org/1367-2630/5/i=1/a=356} {\bibfield
  {journal} {\bibinfo  {journal} {New J. Phys.}\ }\textbf {\bibinfo {volume}
  {5}},\ \bibinfo {pages} {56} (\bibinfo {year} {2003})}\BibitemShut {NoStop}%
\bibitem [{\citenamefont {S\o{}rensen}\ \emph {et~al.}(2005)\citenamefont
  {S\o{}rensen}, \citenamefont {Demler},\ and\ \citenamefont
  {Lukin}}]{Sorensen2005}%
  \BibitemOpen
  \bibfield  {author} {\bibinfo {author} {\bibfnamefont {A.~S.}\ \bibnamefont
  {S\o{}rensen}}, \bibinfo {author} {\bibfnamefont {E.}~\bibnamefont {Demler}},
  \ and\ \bibinfo {author} {\bibfnamefont {M.~D.}\ \bibnamefont {Lukin}},\
  }\href {\doibase 10.1103/PhysRevLett.94.086803} {\bibfield  {journal}
  {\bibinfo  {journal} {Phys. Rev. Lett.}\ }\textbf {\bibinfo {volume} {94}},\
  \bibinfo {pages} {086803} (\bibinfo {year} {2005})}\BibitemShut {NoStop}%
\bibitem [{\citenamefont {Kolovsky}(2011)}]{Kolovsky2011}%
  \BibitemOpen
  \bibfield  {author} {\bibinfo {author} {\bibfnamefont {A.~R.}\ \bibnamefont
  {Kolovsky}},\ }\href {http://stacks.iop.org/0295-5075/93/i=2/a=20003}
  {\bibfield  {journal} {\bibinfo  {journal} {Europhys. Lett.}\ }\textbf
  {\bibinfo {volume} {93}},\ \bibinfo {pages} {20003} (\bibinfo {year}
  {2011})}\BibitemShut {NoStop}%
\bibitem [{\citenamefont {Cooper}(2011)}]{Cooper2011}%
  \BibitemOpen
  \bibfield  {author} {\bibinfo {author} {\bibfnamefont {N.~R.}\ \bibnamefont
  {Cooper}},\ }\href {\doibase 10.1103/PhysRevLett.106.175301} {\bibfield
  {journal} {\bibinfo  {journal} {Phys. Rev. Lett.}\ }\textbf {\bibinfo
  {volume} {106}},\ \bibinfo {pages} {175301} (\bibinfo {year}
  {2011})}\BibitemShut {NoStop}%
\bibitem [{\citenamefont {Struck}\ \emph {et~al.}(2011)\citenamefont {Struck},
  \citenamefont {{\"O}lschl{\"a}ger}, \citenamefont {Le~Targat}, \citenamefont
  {Soltan-Panahi}, \citenamefont {Eckardt}, \citenamefont {Lewenstein},
  \citenamefont {Windpassinger},\ and\ \citenamefont {Sengstock}}]{Struck2011}%
  \BibitemOpen
  \bibfield  {author} {\bibinfo {author} {\bibfnamefont {J.}~\bibnamefont
  {Struck}}, \bibinfo {author} {\bibfnamefont {C.}~\bibnamefont
  {{\"O}lschl{\"a}ger}}, \bibinfo {author} {\bibfnamefont {R.}~\bibnamefont
  {Le~Targat}}, \bibinfo {author} {\bibfnamefont {P.}~\bibnamefont
  {Soltan-Panahi}}, \bibinfo {author} {\bibfnamefont {A.}~\bibnamefont
  {Eckardt}}, \bibinfo {author} {\bibfnamefont {M.}~\bibnamefont {Lewenstein}},
  \bibinfo {author} {\bibfnamefont {P.}~\bibnamefont {Windpassinger}}, \ and\
  \bibinfo {author} {\bibfnamefont {K.}~\bibnamefont {Sengstock}},\ }\href
  {\doibase 10.1126/science.1207239} {\bibfield  {journal} {\bibinfo  {journal}
  {Science}\ }\textbf {\bibinfo {volume} {333}},\ \bibinfo {pages} {996}
  (\bibinfo {year} {2011})}\BibitemShut {NoStop}%
\bibitem [{\citenamefont {Aidelsburger}\ \emph {et~al.}(2011)\citenamefont
  {Aidelsburger}, \citenamefont {Atala}, \citenamefont {Nascimb\`ene},
  \citenamefont {Trotzky}, \citenamefont {Chen},\ and\ \citenamefont
  {Bloch}}]{Aidelsburger2011}%
  \BibitemOpen
  \bibfield  {author} {\bibinfo {author} {\bibfnamefont {M.}~\bibnamefont
  {Aidelsburger}}, \bibinfo {author} {\bibfnamefont {M.}~\bibnamefont {Atala}},
  \bibinfo {author} {\bibfnamefont {S.}~\bibnamefont {Nascimb\`ene}}, \bibinfo
  {author} {\bibfnamefont {S.}~\bibnamefont {Trotzky}}, \bibinfo {author}
  {\bibfnamefont {Y.-A.}\ \bibnamefont {Chen}}, \ and\ \bibinfo {author}
  {\bibfnamefont {I.}~\bibnamefont {Bloch}},\ }\href {\doibase
  10.1103/PhysRevLett.107.255301} {\bibfield  {journal} {\bibinfo  {journal}
  {Phys. Rev. Lett.}\ }\textbf {\bibinfo {volume} {107}},\ \bibinfo {pages}
  {255301} (\bibinfo {year} {2011})}\BibitemShut {NoStop}%
\bibitem [{\citenamefont {Hauke}\ \emph {et~al.}(2012)\citenamefont {Hauke},
  \citenamefont {Tieleman}, \citenamefont {Celi}, \citenamefont
  {\"Olschl\"ager}, \citenamefont {Simonet}, \citenamefont {Struck},
  \citenamefont {Weinberg}, \citenamefont {Windpassinger}, \citenamefont
  {Sengstock}, \citenamefont {Lewenstein},\ and\ \citenamefont
  {Eckardt}}]{Hauke2012}%
  \BibitemOpen
  \bibfield  {author} {\bibinfo {author} {\bibfnamefont {P.}~\bibnamefont
  {Hauke}}, \bibinfo {author} {\bibfnamefont {O.}~\bibnamefont {Tieleman}},
  \bibinfo {author} {\bibfnamefont {A.}~\bibnamefont {Celi}}, \bibinfo {author}
  {\bibfnamefont {C.}~\bibnamefont {\"Olschl\"ager}}, \bibinfo {author}
  {\bibfnamefont {J.}~\bibnamefont {Simonet}}, \bibinfo {author} {\bibfnamefont
  {J.}~\bibnamefont {Struck}}, \bibinfo {author} {\bibfnamefont
  {M.}~\bibnamefont {Weinberg}}, \bibinfo {author} {\bibfnamefont
  {P.}~\bibnamefont {Windpassinger}}, \bibinfo {author} {\bibfnamefont
  {K.}~\bibnamefont {Sengstock}}, \bibinfo {author} {\bibfnamefont
  {M.}~\bibnamefont {Lewenstein}}, \ and\ \bibinfo {author} {\bibfnamefont
  {A.}~\bibnamefont {Eckardt}},\ }\href {\doibase
  10.1103/PhysRevLett.109.145301} {\bibfield  {journal} {\bibinfo  {journal}
  {Phys. Rev. Lett.}\ }\textbf {\bibinfo {volume} {109}},\ \bibinfo {pages}
  {145301} (\bibinfo {year} {2012})}\BibitemShut {NoStop}%
\bibitem [{\citenamefont {Struck}\ \emph {et~al.}(2012)\citenamefont {Struck},
  \citenamefont {\"Olschl\"ager}, \citenamefont {Weinberg}, \citenamefont
  {Hauke}, \citenamefont {Simonet}, \citenamefont {Eckardt}, \citenamefont
  {Lewenstein}, \citenamefont {Sengstock},\ and\ \citenamefont
  {Windpassinger}}]{Struck2012}%
  \BibitemOpen
  \bibfield  {author} {\bibinfo {author} {\bibfnamefont {J.}~\bibnamefont
  {Struck}}, \bibinfo {author} {\bibfnamefont {C.}~\bibnamefont
  {\"Olschl\"ager}}, \bibinfo {author} {\bibfnamefont {M.}~\bibnamefont
  {Weinberg}}, \bibinfo {author} {\bibfnamefont {P.}~\bibnamefont {Hauke}},
  \bibinfo {author} {\bibfnamefont {J.}~\bibnamefont {Simonet}}, \bibinfo
  {author} {\bibfnamefont {A.}~\bibnamefont {Eckardt}}, \bibinfo {author}
  {\bibfnamefont {M.}~\bibnamefont {Lewenstein}}, \bibinfo {author}
  {\bibfnamefont {K.}~\bibnamefont {Sengstock}}, \ and\ \bibinfo {author}
  {\bibfnamefont {P.}~\bibnamefont {Windpassinger}},\ }\href {\doibase
  10.1103/PhysRevLett.108.225304} {\bibfield  {journal} {\bibinfo  {journal}
  {Phys. Rev. Lett.}\ }\textbf {\bibinfo {volume} {108}},\ \bibinfo {pages}
  {225304} (\bibinfo {year} {2012})}\BibitemShut {NoStop}%
\bibitem [{\citenamefont {Miyake}\ \emph {et~al.}(2013)\citenamefont {Miyake},
  \citenamefont {Siviloglou}, \citenamefont {Kennedy}, \citenamefont {Burton},\
  and\ \citenamefont {Ketterle}}]{Miyake2013}%
  \BibitemOpen
  \bibfield  {author} {\bibinfo {author} {\bibfnamefont {H.}~\bibnamefont
  {Miyake}}, \bibinfo {author} {\bibfnamefont {G.~A.}\ \bibnamefont
  {Siviloglou}}, \bibinfo {author} {\bibfnamefont {C.~J.}\ \bibnamefont
  {Kennedy}}, \bibinfo {author} {\bibfnamefont {W.~C.}\ \bibnamefont {Burton}},
  \ and\ \bibinfo {author} {\bibfnamefont {W.}~\bibnamefont {Ketterle}},\
  }\href {\doibase 10.1103/PhysRevLett.111.185302} {\bibfield  {journal}
  {\bibinfo  {journal} {Phys. Rev. Lett.}\ }\textbf {\bibinfo {volume} {111}},\
  \bibinfo {pages} {185302} (\bibinfo {year} {2013})}\BibitemShut {NoStop}%
\bibitem [{\citenamefont {Aidelsburger}\ \emph {et~al.}(2013)\citenamefont
  {Aidelsburger}, \citenamefont {Atala}, \citenamefont {Lohse}, \citenamefont
  {Barreiro}, \citenamefont {Paredes},\ and\ \citenamefont
  {Bloch}}]{Aidelsburger2013}%
  \BibitemOpen
  \bibfield  {author} {\bibinfo {author} {\bibfnamefont {M.}~\bibnamefont
  {Aidelsburger}}, \bibinfo {author} {\bibfnamefont {M.}~\bibnamefont {Atala}},
  \bibinfo {author} {\bibfnamefont {M.}~\bibnamefont {Lohse}}, \bibinfo
  {author} {\bibfnamefont {J.~T.}\ \bibnamefont {Barreiro}}, \bibinfo {author}
  {\bibfnamefont {B.}~\bibnamefont {Paredes}}, \ and\ \bibinfo {author}
  {\bibfnamefont {I.}~\bibnamefont {Bloch}},\ }\href {\doibase
  10.1103/PhysRevLett.111.185301} {\bibfield  {journal} {\bibinfo  {journal}
  {Phys. Rev. Lett.}\ }\textbf {\bibinfo {volume} {111}},\ \bibinfo {pages}
  {185301} (\bibinfo {year} {2013})}\BibitemShut {NoStop}%
\bibitem [{\citenamefont {Atala}\ \emph {et~al.}(2014)\citenamefont {Atala},
  \citenamefont {Aidelsburger}, \citenamefont {Lohse}, \citenamefont
  {Barreiro}, \citenamefont {Paredes},\ and\ \citenamefont {Bloch}}]{atala14}%
  \BibitemOpen
  \bibfield  {author} {\bibinfo {author} {\bibfnamefont {M.}~\bibnamefont
  {Atala}}, \bibinfo {author} {\bibfnamefont {M.}~\bibnamefont {Aidelsburger}},
  \bibinfo {author} {\bibfnamefont {M.}~\bibnamefont {Lohse}}, \bibinfo
  {author} {\bibfnamefont {J.~T.}\ \bibnamefont {Barreiro}}, \bibinfo {author}
  {\bibfnamefont {B.}~\bibnamefont {Paredes}}, \ and\ \bibinfo {author}
  {\bibfnamefont {I.}~\bibnamefont {Bloch}},\ }\href@noop {} {\bibfield
  {journal} {\bibinfo  {journal} {Nat. Phys.}\ }\textbf {\bibinfo {volume}
  {10}},\ \bibinfo {pages} {588} (\bibinfo {year} {2014})}\BibitemShut
  {NoStop}%
\bibitem [{\citenamefont {Dalibard}\ \emph {et~al.}(2011)\citenamefont
  {Dalibard}, \citenamefont {Gerbier}, \citenamefont
  {Juzeli\ifmmode~\bar{u}\else \={u}\fi{}nas},\ and\ \citenamefont
  {\"Ohberg}}]{Dalibard2011}%
  \BibitemOpen
  \bibfield  {author} {\bibinfo {author} {\bibfnamefont {J.}~\bibnamefont
  {Dalibard}}, \bibinfo {author} {\bibfnamefont {F.}~\bibnamefont {Gerbier}},
  \bibinfo {author} {\bibfnamefont {G.}~\bibnamefont
  {Juzeli\ifmmode~\bar{u}\else \={u}\fi{}nas}}, \ and\ \bibinfo {author}
  {\bibfnamefont {P.}~\bibnamefont {\"Ohberg}},\ }\href {\doibase
  10.1103/RevModPhys.83.1523} {\bibfield  {journal} {\bibinfo  {journal} {Rev.
  Mod. Phys.}\ }\textbf {\bibinfo {volume} {83}},\ \bibinfo {pages} {1523}
  (\bibinfo {year} {2011})}\BibitemShut {NoStop}%
\bibitem [{\citenamefont {Goldman}\ \emph {et~al.}(2014)\citenamefont
  {Goldman}, \citenamefont {Juzeli{\~u}nas}, \citenamefont {{\"O}hberg},\ and\
  \citenamefont {Spielman}}]{Goldman2014}%
  \BibitemOpen
  \bibfield  {author} {\bibinfo {author} {\bibfnamefont {N.}~\bibnamefont
  {Goldman}}, \bibinfo {author} {\bibfnamefont {G.}~\bibnamefont
  {Juzeli{\~u}nas}}, \bibinfo {author} {\bibfnamefont {P.}~\bibnamefont
  {{\"O}hberg}}, \ and\ \bibinfo {author} {\bibfnamefont {I.~B.}\ \bibnamefont
  {Spielman}},\ }\href {http://stacks.iop.org/0034-4885/77/i=12/a=126401}
  {\bibfield  {journal} {\bibinfo  {journal} {Rep. Prog. Phys.}\ }\textbf
  {\bibinfo {volume} {77}},\ \bibinfo {pages} {126401} (\bibinfo {year}
  {2014})}\BibitemShut {NoStop}%
\bibitem [{\citenamefont {Zhai}(2012)}]{Zhai12}%
  \BibitemOpen
  \bibfield  {author} {\bibinfo {author} {\bibfnamefont {H.}~\bibnamefont
  {Zhai}},\ }\href {\doibase 10.1142/S0217979212300010} {\bibfield  {journal}
  {\bibinfo  {journal} {Int. J. Mod. Phys. B}\ }\textbf {\bibinfo {volume}
  {26}},\ \bibinfo {pages} {1230001} (\bibinfo {year} {2012})}\BibitemShut
  {NoStop}%
\bibitem [{\citenamefont {Zhai}(2015)}]{Zhai15}%
  \BibitemOpen
  \bibfield  {author} {\bibinfo {author} {\bibfnamefont {H.}~\bibnamefont
  {Zhai}},\ }\href {http://stacks.iop.org/0034-4885/78/i=2/a=026001} {\bibfield
   {journal} {\bibinfo  {journal} {Rep. Prog. Phys.}\ }\textbf {\bibinfo
  {volume} {78}},\ \bibinfo {pages} {026001} (\bibinfo {year}
  {2015})}\BibitemShut {NoStop}%
\bibitem [{\citenamefont {Schine}\ \emph {et~al.}(2016)\citenamefont {Schine},
  \citenamefont {Ryou}, \citenamefont {Gromov}, \citenamefont {Sommer},\ and\
  \citenamefont {Simon}}]{Schine2016}%
  \BibitemOpen
  \bibfield  {author} {\bibinfo {author} {\bibfnamefont {N.}~\bibnamefont
  {Schine}}, \bibinfo {author} {\bibfnamefont {A.}~\bibnamefont {Ryou}},
  \bibinfo {author} {\bibfnamefont {A.}~\bibnamefont {Gromov}}, \bibinfo
  {author} {\bibfnamefont {A.}~\bibnamefont {Sommer}}, \ and\ \bibinfo {author}
  {\bibfnamefont {J.}~\bibnamefont {Simon}},\ }\href
  {http://dx.doi.org/10.1038/nature17943} {\bibfield  {journal} {\bibinfo
  {journal} {Nature (London)}\ }\textbf {\bibinfo {volume} {534}},\ \bibinfo
  {pages} {671} (\bibinfo {year} {2016})}\BibitemShut {NoStop}%
\bibitem [{\citenamefont {Fang}\ \emph {et~al.}(2012)\citenamefont {Fang},
  \citenamefont {Yu},\ and\ \citenamefont {Fan}}]{Fang2012}%
  \BibitemOpen
  \bibfield  {author} {\bibinfo {author} {\bibfnamefont {K.}~\bibnamefont
  {Fang}}, \bibinfo {author} {\bibfnamefont {Z.}~\bibnamefont {Yu}}, \ and\
  \bibinfo {author} {\bibfnamefont {S.}~\bibnamefont {Fan}},\ }\href {\doibase
  10.1038/nphoton.2012.236} {\bibfield  {journal} {\bibinfo  {journal} {Nat.
  Photon.}\ }\textbf {\bibinfo {volume} {6}},\ \bibinfo {pages} {782} (\bibinfo
  {year} {2012})}\BibitemShut {NoStop}%
\bibitem [{\citenamefont {Hafezi}\ \emph {et~al.}(2013)\citenamefont {Hafezi},
  \citenamefont {Mittal}, \citenamefont {Fan}, \citenamefont {Migdall},\ and\
  \citenamefont {Taylor}}]{Hafezi2013}%
  \BibitemOpen
  \bibfield  {author} {\bibinfo {author} {\bibfnamefont {M.}~\bibnamefont
  {Hafezi}}, \bibinfo {author} {\bibfnamefont {S.}~\bibnamefont {Mittal}},
  \bibinfo {author} {\bibfnamefont {J.}~\bibnamefont {Fan}}, \bibinfo {author}
  {\bibfnamefont {A.}~\bibnamefont {Migdall}}, \ and\ \bibinfo {author}
  {\bibfnamefont {J.~M.}\ \bibnamefont {Taylor}},\ }\href
  {http://dx.doi.org/10.1038/nphoton.2013.274} {\bibfield  {journal} {\bibinfo
  {journal} {Nat. Photon.}\ }\textbf {\bibinfo {volume} {7}},\ \bibinfo {pages}
  {1001} (\bibinfo {year} {2013})}\BibitemShut {NoStop}%
\bibitem [{\citenamefont {Dub{\v c}ek}\ \emph {et~al.}(2015)\citenamefont
  {Dub{\v c}ek}, \citenamefont {Lelas}, \citenamefont {Juki{\'c}},
  \citenamefont {Pezer}, \citenamefont {Solja{\v c}i{\'c}},\ and\ \citenamefont
  {Buljan}}]{Dubcek2015}%
  \BibitemOpen
  \bibfield  {author} {\bibinfo {author} {\bibfnamefont {T.}~\bibnamefont
  {Dub{\v c}ek}}, \bibinfo {author} {\bibfnamefont {K.}~\bibnamefont {Lelas}},
  \bibinfo {author} {\bibfnamefont {D.}~\bibnamefont {Juki{\'c}}}, \bibinfo
  {author} {\bibfnamefont {R.}~\bibnamefont {Pezer}}, \bibinfo {author}
  {\bibfnamefont {M.}~\bibnamefont {Solja{\v c}i{\'c}}}, \ and\ \bibinfo
  {author} {\bibfnamefont {H.}~\bibnamefont {Buljan}},\ }\href
  {http://stacks.iop.org/1367-2630/17/i=12/a=125002} {\bibfield  {journal}
  {\bibinfo  {journal} {New J. Phys.}\ }\textbf {\bibinfo {volume} {17}},\
  \bibinfo {pages} {125002} (\bibinfo {year} {2015})}\BibitemShut {NoStop}%
\bibitem [{\citenamefont {Hess}\ and\ \citenamefont {Fairbank}(1967)}]{Hess67}%
  \BibitemOpen
  \bibfield  {author} {\bibinfo {author} {\bibfnamefont {G.~B.}\ \bibnamefont
  {Hess}}\ and\ \bibinfo {author} {\bibfnamefont {W.~M.}\ \bibnamefont
  {Fairbank}},\ }\href@noop {} {\bibfield  {journal} {\bibinfo  {journal}
  {Phys. Rev. Lett.}\ }\textbf {\bibinfo {volume} {19}},\ \bibinfo {pages}
  {216} (\bibinfo {year} {1967})}\BibitemShut {NoStop}%
\bibitem [{\citenamefont {{Walter}}\ and\ \citenamefont
  {{Marquardt}}(2015)}]{Walter2015}%
  \BibitemOpen
  \bibfield  {author} {\bibinfo {author} {\bibfnamefont {S.}~\bibnamefont
  {{Walter}}}\ and\ \bibinfo {author} {\bibfnamefont {F.}~\bibnamefont
  {{Marquardt}}},\ }\href@noop {} {\enquote {\bibinfo {title} {{Dynamical Gauge
  Fields in Optomechanics}},}\ } (\bibinfo {year} {2015}),\ \Eprint
  {http://arxiv.org/abs/1510.06754} {arXiv:1510.06754} \BibitemShut {NoStop}%
\bibitem [{\citenamefont {{Zheng}}\ and\ \citenamefont
  {{Cooper}}(2016)}]{Zheng2016}%
  \BibitemOpen
  \bibfield  {author} {\bibinfo {author} {\bibfnamefont {W.}~\bibnamefont
  {{Zheng}}}\ and\ \bibinfo {author} {\bibfnamefont {N.~R.}\ \bibnamefont
  {{Cooper}}},\ }\href@noop {} {\enquote {\bibinfo {title} {{Superradiance
  induced particle flow via dynamical gauge coupling}},}\ } (\bibinfo {year}
  {2016}),\ \Eprint {http://arxiv.org/abs/1604.06630} {arXiv:1604.06630}
  \BibitemShut {NoStop}%
\bibitem [{\citenamefont {Domokos}\ and\ \citenamefont
  {Ritsch}(2002)}]{Domokos2002}%
  \BibitemOpen
  \bibfield  {author} {\bibinfo {author} {\bibfnamefont {P.}~\bibnamefont
  {Domokos}}\ and\ \bibinfo {author} {\bibfnamefont {H.}~\bibnamefont
  {Ritsch}},\ }\href {\doibase 10.1103/PhysRevLett.89.253003} {\bibfield
  {journal} {\bibinfo  {journal} {Phys. Rev. Lett.}\ }\textbf {\bibinfo
  {volume} {89}},\ \bibinfo {pages} {253003} (\bibinfo {year}
  {2002})}\BibitemShut {NoStop}%
\bibitem [{\citenamefont {Black}\ \emph {et~al.}(2003)\citenamefont {Black},
  \citenamefont {Chan},\ and\ \citenamefont {Vuleti\ifmmode~\acute{c}\else
  \'{c}\fi{}}}]{Black2003}%
  \BibitemOpen
  \bibfield  {author} {\bibinfo {author} {\bibfnamefont {A.~T.}\ \bibnamefont
  {Black}}, \bibinfo {author} {\bibfnamefont {H.~W.}\ \bibnamefont {Chan}}, \
  and\ \bibinfo {author} {\bibfnamefont {V.}~\bibnamefont
  {Vuleti\ifmmode~\acute{c}\else \'{c}\fi{}}},\ }\href {\doibase
  10.1103/PhysRevLett.91.203001} {\bibfield  {journal} {\bibinfo  {journal}
  {Phys. Rev. Lett.}\ }\textbf {\bibinfo {volume} {91}},\ \bibinfo {pages}
  {203001} (\bibinfo {year} {2003})}\BibitemShut {NoStop}%
\bibitem [{\citenamefont {Baumann}\ \emph {et~al.}(2010)\citenamefont
  {Baumann}, \citenamefont {Guerlin}, \citenamefont {Brennecke},\ and\
  \citenamefont {Esslinger}}]{Baumann2010}%
  \BibitemOpen
  \bibfield  {author} {\bibinfo {author} {\bibfnamefont {K.}~\bibnamefont
  {Baumann}}, \bibinfo {author} {\bibfnamefont {C.}~\bibnamefont {Guerlin}},
  \bibinfo {author} {\bibfnamefont {F.}~\bibnamefont {Brennecke}}, \ and\
  \bibinfo {author} {\bibfnamefont {T.}~\bibnamefont {Esslinger}},\ }\href
  {\doibase 10.1038/nature09009} {\bibfield  {journal} {\bibinfo  {journal}
  {Nature (London)}\ }\textbf {\bibinfo {volume} {464}},\ \bibinfo {pages}
  {1301} (\bibinfo {year} {2010})}\BibitemShut {NoStop}%
\bibitem [{\citenamefont {Deng}\ \emph {et~al.}(2014)\citenamefont {Deng},
  \citenamefont {Cheng}, \citenamefont {Jing},\ and\ \citenamefont
  {Yi}}]{deng14}%
  \BibitemOpen
  \bibfield  {author} {\bibinfo {author} {\bibfnamefont {Y.}~\bibnamefont
  {Deng}}, \bibinfo {author} {\bibfnamefont {J.}~\bibnamefont {Cheng}},
  \bibinfo {author} {\bibfnamefont {H.}~\bibnamefont {Jing}}, \ and\ \bibinfo
  {author} {\bibfnamefont {S.}~\bibnamefont {Yi}},\ }\href {\doibase
  10.1103/PhysRevLett.112.143007} {\bibfield  {journal} {\bibinfo  {journal}
  {Phys. Rev. Lett.}\ }\textbf {\bibinfo {volume} {112}},\ \bibinfo {pages}
  {143007} (\bibinfo {year} {2014})}\BibitemShut {NoStop}%
\bibitem [{\citenamefont {Dong}\ \emph {et~al.}(2014)\citenamefont {Dong},
  \citenamefont {Zhou}, \citenamefont {Wu}, \citenamefont {Ramachandhran},\
  and\ \citenamefont {Pu}}]{dong14}%
  \BibitemOpen
  \bibfield  {author} {\bibinfo {author} {\bibfnamefont {L.}~\bibnamefont
  {Dong}}, \bibinfo {author} {\bibfnamefont {L.}~\bibnamefont {Zhou}}, \bibinfo
  {author} {\bibfnamefont {B.}~\bibnamefont {Wu}}, \bibinfo {author}
  {\bibfnamefont {B.}~\bibnamefont {Ramachandhran}}, \ and\ \bibinfo {author}
  {\bibfnamefont {H.}~\bibnamefont {Pu}},\ }\href {\doibase
  10.1103/PhysRevA.89.011602} {\bibfield  {journal} {\bibinfo  {journal} {Phys.
  Rev. A}\ }\textbf {\bibinfo {volume} {89}},\ \bibinfo {pages} {011602}
  (\bibinfo {year} {2014})}\BibitemShut {NoStop}%
\bibitem [{\citenamefont {Padhi}\ and\ \citenamefont {Ghosh}(2014)}]{padhi14}%
  \BibitemOpen
  \bibfield  {author} {\bibinfo {author} {\bibfnamefont {B.}~\bibnamefont
  {Padhi}}\ and\ \bibinfo {author} {\bibfnamefont {S.}~\bibnamefont {Ghosh}},\
  }\href {\doibase 10.1103/PhysRevA.90.023627} {\bibfield  {journal} {\bibinfo
  {journal} {Phys. Rev. A}\ }\textbf {\bibinfo {volume} {90}},\ \bibinfo
  {pages} {023627} (\bibinfo {year} {2014})}\BibitemShut {NoStop}%
\bibitem [{\citenamefont {Pan}\ \emph {et~al.}(2015)\citenamefont {Pan},
  \citenamefont {Liu}, \citenamefont {Zhang}, \citenamefont {Yi},\ and\
  \citenamefont {Guo}}]{pan15}%
  \BibitemOpen
  \bibfield  {author} {\bibinfo {author} {\bibfnamefont {J.-S.}\ \bibnamefont
  {Pan}}, \bibinfo {author} {\bibfnamefont {X.-J.}\ \bibnamefont {Liu}},
  \bibinfo {author} {\bibfnamefont {W.}~\bibnamefont {Zhang}}, \bibinfo
  {author} {\bibfnamefont {W.}~\bibnamefont {Yi}}, \ and\ \bibinfo {author}
  {\bibfnamefont {G.-C.}\ \bibnamefont {Guo}},\ }\href {\doibase
  10.1103/PhysRevLett.115.045303} {\bibfield  {journal} {\bibinfo  {journal}
  {Phys. Rev. Lett.}\ }\textbf {\bibinfo {volume} {115}},\ \bibinfo {pages}
  {045303} (\bibinfo {year} {2015})}\BibitemShut {NoStop}%
\bibitem [{\citenamefont {Dong}\ \emph {et~al.}(2015)\citenamefont {Dong},
  \citenamefont {Zhu},\ and\ \citenamefont {Pu}}]{Dong2015}%
  \BibitemOpen
  \bibfield  {author} {\bibinfo {author} {\bibfnamefont {L.}~\bibnamefont
  {Dong}}, \bibinfo {author} {\bibfnamefont {C.}~\bibnamefont {Zhu}}, \ and\
  \bibinfo {author} {\bibfnamefont {H.}~\bibnamefont {Pu}},\ }\href {\doibase
  10.3390/atoms3020182} {\bibfield  {journal} {\bibinfo  {journal} {Atoms}\
  }\textbf {\bibinfo {volume} {3}},\ \bibinfo {pages} {182} (\bibinfo {year}
  {2015})}\BibitemShut {NoStop}%
\bibitem [{\citenamefont {Kollath}\ \emph {et~al.}(2016)\citenamefont
  {Kollath}, \citenamefont {Sheikhan}, \citenamefont {Wolff},\ and\
  \citenamefont {Brennecke}}]{Kollath2016}%
  \BibitemOpen
  \bibfield  {author} {\bibinfo {author} {\bibfnamefont {C.}~\bibnamefont
  {Kollath}}, \bibinfo {author} {\bibfnamefont {A.}~\bibnamefont {Sheikhan}},
  \bibinfo {author} {\bibfnamefont {S.}~\bibnamefont {Wolff}}, \ and\ \bibinfo
  {author} {\bibfnamefont {F.}~\bibnamefont {Brennecke}},\ }\href {\doibase
  10.1103/PhysRevLett.116.060401} {\bibfield  {journal} {\bibinfo  {journal}
  {Phys. Rev. Lett.}\ }\textbf {\bibinfo {volume} {116}},\ \bibinfo {pages}
  {060401} (\bibinfo {year} {2016})}\BibitemShut {NoStop}%
\bibitem [{\citenamefont {Sheikhan}\ \emph {et~al.}(2016)\citenamefont
  {Sheikhan}, \citenamefont {Brennecke},\ and\ \citenamefont
  {Kollath}}]{Sheikhan2016}%
  \BibitemOpen
  \bibfield  {author} {\bibinfo {author} {\bibfnamefont {A.}~\bibnamefont
  {Sheikhan}}, \bibinfo {author} {\bibfnamefont {F.}~\bibnamefont {Brennecke}},
  \ and\ \bibinfo {author} {\bibfnamefont {C.}~\bibnamefont {Kollath}},\ }\href
  {\doibase 10.1103/PhysRevA.93.043609} {\bibfield  {journal} {\bibinfo
  {journal} {Phys. Rev. A}\ }\textbf {\bibinfo {volume} {93}},\ \bibinfo
  {pages} {043609} (\bibinfo {year} {2016})}\BibitemShut {NoStop}%
\bibitem [{\citenamefont {Gopalakrishnan}\ \emph {et~al.}(2009)\citenamefont
  {Gopalakrishnan}, \citenamefont {Lev},\ and\ \citenamefont
  {Goldbart}}]{Gopalakrishnan2009}%
  \BibitemOpen
  \bibfield  {author} {\bibinfo {author} {\bibfnamefont {S.}~\bibnamefont
  {Gopalakrishnan}}, \bibinfo {author} {\bibfnamefont {B.~L.}\ \bibnamefont
  {Lev}}, \ and\ \bibinfo {author} {\bibfnamefont {P.~M.}\ \bibnamefont
  {Goldbart}},\ }\href {\doibase 10.1038/nphys1403} {\bibfield  {journal}
  {\bibinfo  {journal} {Nat. Phys.}\ }\textbf {\bibinfo {volume} {5}},\
  \bibinfo {pages} {845} (\bibinfo {year} {2009})}\BibitemShut {NoStop}%
\bibitem [{\citenamefont {Talbot}(1836)}]{talbot36}%
  \BibitemOpen
  \bibfield  {author} {\bibinfo {author} {\bibfnamefont {H.}~\bibnamefont
  {Talbot}},\ }\href {\doibase 10.1080/14786443608649032} {\bibfield  {journal}
  {\bibinfo  {journal} {Philos. Mag. Series 3}\ }\textbf {\bibinfo {volume}
  {9}},\ \bibinfo {pages} {401} (\bibinfo {year} {1836})}\BibitemShut {NoStop}%
\bibitem [{\citenamefont {Ackemann}\ and\ \citenamefont
  {Lange}(2001)}]{Ackemann2001}%
  \BibitemOpen
  \bibfield  {author} {\bibinfo {author} {\bibfnamefont {T.}~\bibnamefont
  {Ackemann}}\ and\ \bibinfo {author} {\bibfnamefont {W.}~\bibnamefont
  {Lange}},\ }\href {\doibase 10.1007/s003400000518} {\bibfield  {journal}
  {\bibinfo  {journal} {Appl. Phys. B}\ }\textbf {\bibinfo {volume} {72}},\
  \bibinfo {pages} {21} (\bibinfo {year} {2001})}\BibitemShut {NoStop}%
\bibitem [{\citenamefont {Labeyrie}\ \emph {et~al.}(2014)\citenamefont
  {Labeyrie}, \citenamefont {Tesio}, \citenamefont {Gomes}, \citenamefont
  {Oppo}, \citenamefont {Firth}, \citenamefont {Robb}, \citenamefont {Arnold},
  \citenamefont {Kaiser},\ and\ \citenamefont {Ackemann}}]{Labeyrie2014}%
  \BibitemOpen
  \bibfield  {author} {\bibinfo {author} {\bibfnamefont {G.}~\bibnamefont
  {Labeyrie}}, \bibinfo {author} {\bibfnamefont {E.}~\bibnamefont {Tesio}},
  \bibinfo {author} {\bibfnamefont {P.~M.}\ \bibnamefont {Gomes}}, \bibinfo
  {author} {\bibfnamefont {G.-L.}\ \bibnamefont {Oppo}}, \bibinfo {author}
  {\bibfnamefont {W.~J.}\ \bibnamefont {Firth}}, \bibinfo {author}
  {\bibfnamefont {G.~R.~M.}\ \bibnamefont {Robb}}, \bibinfo {author}
  {\bibfnamefont {A.~S.}\ \bibnamefont {Arnold}}, \bibinfo {author}
  {\bibfnamefont {R.}~\bibnamefont {Kaiser}}, \ and\ \bibinfo {author}
  {\bibfnamefont {T.}~\bibnamefont {Ackemann}},\ }\href {\doibase
  10.1038/nphoton.2014.52} {\bibfield  {journal} {\bibinfo  {journal} {Nat.
  Photon.}\ }\textbf {\bibinfo {volume} {8}},\ \bibinfo {pages} {321} (\bibinfo
  {year} {2014})}\BibitemShut {NoStop}%
\bibitem [{\citenamefont {Diver}\ \emph {et~al.}(2014)\citenamefont {Diver},
  \citenamefont {Robb},\ and\ \citenamefont {Oppo}}]{Diver2014}%
  \BibitemOpen
  \bibfield  {author} {\bibinfo {author} {\bibfnamefont {M.}~\bibnamefont
  {Diver}}, \bibinfo {author} {\bibfnamefont {G.~R.~M.}\ \bibnamefont {Robb}},
  \ and\ \bibinfo {author} {\bibfnamefont {G.-L.}\ \bibnamefont {Oppo}},\
  }\href {\doibase 10.1103/PhysRevA.89.033602} {\bibfield  {journal} {\bibinfo
  {journal} {Phys. Rev. A}\ }\textbf {\bibinfo {volume} {89}},\ \bibinfo
  {pages} {033602} (\bibinfo {year} {2014})}\BibitemShut {NoStop}%
\bibitem [{\citenamefont {Robb}\ \emph {et~al.}(2015)\citenamefont {Robb},
  \citenamefont {Tesio}, \citenamefont {Oppo}, \citenamefont {Firth},
  \citenamefont {Ackemann},\ and\ \citenamefont {Bonifacio}}]{Robb2015}%
  \BibitemOpen
  \bibfield  {author} {\bibinfo {author} {\bibfnamefont {G.~R.~M.}\
  \bibnamefont {Robb}}, \bibinfo {author} {\bibfnamefont {E.}~\bibnamefont
  {Tesio}}, \bibinfo {author} {\bibfnamefont {G.-L.}\ \bibnamefont {Oppo}},
  \bibinfo {author} {\bibfnamefont {W.~J.}\ \bibnamefont {Firth}}, \bibinfo
  {author} {\bibfnamefont {T.}~\bibnamefont {Ackemann}}, \ and\ \bibinfo
  {author} {\bibfnamefont {R.}~\bibnamefont {Bonifacio}},\ }\href {\doibase
  10.1103/PhysRevLett.114.173903} {\bibfield  {journal} {\bibinfo  {journal}
  {Phys. Rev. Lett.}\ }\textbf {\bibinfo {volume} {114}},\ \bibinfo {pages}
  {173903} (\bibinfo {year} {2015})}\BibitemShut {NoStop}%
\bibitem [{\citenamefont {Gopalakrishnan}\ \emph {et~al.}(2010)\citenamefont
  {Gopalakrishnan}, \citenamefont {Lev},\ and\ \citenamefont
  {Goldbart}}]{Gopalakrishnan2010}%
  \BibitemOpen
  \bibfield  {author} {\bibinfo {author} {\bibfnamefont {S.}~\bibnamefont
  {Gopalakrishnan}}, \bibinfo {author} {\bibfnamefont {B.}~\bibnamefont {Lev}},
  \ and\ \bibinfo {author} {\bibfnamefont {P.}~\bibnamefont {Goldbart}},\
  }\href {\doibase 10.1103/PhysRevA.82.043612} {\bibfield  {journal} {\bibinfo
  {journal} {Phys. Rev. A}\ }\textbf {\bibinfo {volume} {82}},\ \bibinfo
  {pages} {34} (\bibinfo {year} {2010})}\BibitemShut {NoStop}%
\bibitem [{\citenamefont {Wickenbrock}\ \emph {et~al.}(2013)\citenamefont
  {Wickenbrock}, \citenamefont {Hemmerling}, \citenamefont {Robb},
  \citenamefont {Emary},\ and\ \citenamefont {Renzoni}}]{Wickenbrock2013}%
  \BibitemOpen
  \bibfield  {author} {\bibinfo {author} {\bibfnamefont {A.}~\bibnamefont
  {Wickenbrock}}, \bibinfo {author} {\bibfnamefont {M.}~\bibnamefont
  {Hemmerling}}, \bibinfo {author} {\bibfnamefont {G.~R.~M.}\ \bibnamefont
  {Robb}}, \bibinfo {author} {\bibfnamefont {C.}~\bibnamefont {Emary}}, \ and\
  \bibinfo {author} {\bibfnamefont {F.}~\bibnamefont {Renzoni}},\ }\href
  {\doibase 10.1103/PhysRevA.87.043817} {\bibfield  {journal} {\bibinfo
  {journal} {Phys. Rev. A}\ }\textbf {\bibinfo {volume} {87}},\ \bibinfo
  {pages} {043817} (\bibinfo {year} {2013})}\BibitemShut {NoStop}%
\bibitem [{\citenamefont {Koll{\'{a}}r}\ \emph {et~al.}(2015)\citenamefont
  {Koll{\'{a}}r}, \citenamefont {Papageorge}, \citenamefont {Baumann},
  \citenamefont {Armen},\ and\ \citenamefont {Lev}}]{Kollar2015}%
  \BibitemOpen
  \bibfield  {author} {\bibinfo {author} {\bibfnamefont {A.~J.}\ \bibnamefont
  {Koll{\'{a}}r}}, \bibinfo {author} {\bibfnamefont {A.~T.}\ \bibnamefont
  {Papageorge}}, \bibinfo {author} {\bibfnamefont {K.}~\bibnamefont {Baumann}},
  \bibinfo {author} {\bibfnamefont {M.~A.}\ \bibnamefont {Armen}}, \ and\
  \bibinfo {author} {\bibfnamefont {B.~L.}\ \bibnamefont {Lev}},\ }\href
  {\doibase 10.1088/1367-2630/17/4/043012} {\bibfield  {journal} {\bibinfo
  {journal} {New J. Phys.}\ }\textbf {\bibinfo {volume} {17}},\ \bibinfo
  {pages} {043012} (\bibinfo {year} {2015})}\BibitemShut {NoStop}%
\bibitem [{\citenamefont {{Koll{\'a}r}}\ \emph {et~al.}(2016)\citenamefont
  {{Koll{\'a}r}}, \citenamefont {{Papageorge}}, \citenamefont {{Vaidya}},
  \citenamefont {{Guo}}, \citenamefont {{Keeling}},\ and\ \citenamefont
  {{Lev}}}]{Kollar2016}%
  \BibitemOpen
  \bibfield  {author} {\bibinfo {author} {\bibfnamefont {A.~J.}\ \bibnamefont
  {{Koll{\'a}r}}}, \bibinfo {author} {\bibfnamefont {A.~T.}\ \bibnamefont
  {{Papageorge}}}, \bibinfo {author} {\bibfnamefont {V.~D.}\ \bibnamefont
  {{Vaidya}}}, \bibinfo {author} {\bibfnamefont {Y.}~\bibnamefont {{Guo}}},
  \bibinfo {author} {\bibfnamefont {J.}~\bibnamefont {{Keeling}}}, \ and\
  \bibinfo {author} {\bibfnamefont {B.~L.}\ \bibnamefont {{Lev}}},\ }\href@noop
  {} {\enquote {\bibinfo {title} {{Supermode-Density-Wave-Polariton
  Condensation}},}\ } (\bibinfo {year} {2016}),\ \Eprint
  {http://arxiv.org/abs/1606.04127} {arXiv:1606.04127} \BibitemShut {NoStop}%
\bibitem [{\citenamefont {Tewari}\ \emph {et~al.}(2006)\citenamefont {Tewari},
  \citenamefont {Scarola}, \citenamefont {Senthil},\ and\ \citenamefont
  {Sarma}}]{Tewari2006}%
  \BibitemOpen
  \bibfield  {author} {\bibinfo {author} {\bibfnamefont {S.}~\bibnamefont
  {Tewari}}, \bibinfo {author} {\bibfnamefont {V.~W.}\ \bibnamefont {Scarola}},
  \bibinfo {author} {\bibfnamefont {T.}~\bibnamefont {Senthil}}, \ and\
  \bibinfo {author} {\bibfnamefont {S.~D.}\ \bibnamefont {Sarma}},\ }\href
  {\doibase 10.1103/PhysRevLett.97.200401} {\bibfield  {journal} {\bibinfo
  {journal} {Phys. Rev. Lett.}\ }\textbf {\bibinfo {volume} {97}},\ \bibinfo
  {pages} {200401} (\bibinfo {year} {2006})}\BibitemShut {NoStop}%
\bibitem [{\citenamefont {Cirac}\ \emph {et~al.}(2010)\citenamefont {Cirac},
  \citenamefont {Maraner},\ and\ \citenamefont {Pachos}}]{Cirac2010}%
  \BibitemOpen
  \bibfield  {author} {\bibinfo {author} {\bibfnamefont {J.~I.}\ \bibnamefont
  {Cirac}}, \bibinfo {author} {\bibfnamefont {P.}~\bibnamefont {Maraner}}, \
  and\ \bibinfo {author} {\bibfnamefont {J.~K.}\ \bibnamefont {Pachos}},\
  }\href {\doibase 10.1103/PhysRevLett.105.190403} {\bibfield  {journal}
  {\bibinfo  {journal} {Phys. Rev. Lett.}\ }\textbf {\bibinfo {volume} {105}},\
  \bibinfo {pages} {190403} (\bibinfo {year} {2010})}\BibitemShut {NoStop}%
\bibitem [{\citenamefont {Zohar}\ and\ \citenamefont
  {Reznik}(2011)}]{Zohar2011}%
  \BibitemOpen
  \bibfield  {author} {\bibinfo {author} {\bibfnamefont {E.}~\bibnamefont
  {Zohar}}\ and\ \bibinfo {author} {\bibfnamefont {B.}~\bibnamefont {Reznik}},\
  }\href {\doibase 10.1103/PhysRevLett.107.275301} {\bibfield  {journal}
  {\bibinfo  {journal} {Phys. Rev. Lett.}\ }\textbf {\bibinfo {volume} {107}},\
  \bibinfo {pages} {275301} (\bibinfo {year} {2011})}\BibitemShut {NoStop}%
\bibitem [{\citenamefont {Zohar}\ \emph {et~al.}(2012)\citenamefont {Zohar},
  \citenamefont {Cirac},\ and\ \citenamefont {Reznik}}]{Zohar2012}%
  \BibitemOpen
  \bibfield  {author} {\bibinfo {author} {\bibfnamefont {E.}~\bibnamefont
  {Zohar}}, \bibinfo {author} {\bibfnamefont {J.~I.}\ \bibnamefont {Cirac}}, \
  and\ \bibinfo {author} {\bibfnamefont {B.}~\bibnamefont {Reznik}},\ }\href
  {\doibase 10.1103/PhysRevLett.109.125302} {\bibfield  {journal} {\bibinfo
  {journal} {Phys. Rev. Lett.}\ }\textbf {\bibinfo {volume} {109}},\ \bibinfo
  {pages} {125302} (\bibinfo {year} {2012})}\BibitemShut {NoStop}%
\bibitem [{\citenamefont {Banerjee}\ \emph {et~al.}(2012)\citenamefont
  {Banerjee}, \citenamefont {Dalmonte}, \citenamefont {M\"uller}, \citenamefont
  {Rico}, \citenamefont {Stebler}, \citenamefont {Wiese},\ and\ \citenamefont
  {Zoller}}]{Banerjee2012}%
  \BibitemOpen
  \bibfield  {author} {\bibinfo {author} {\bibfnamefont {D.}~\bibnamefont
  {Banerjee}}, \bibinfo {author} {\bibfnamefont {M.}~\bibnamefont {Dalmonte}},
  \bibinfo {author} {\bibfnamefont {M.}~\bibnamefont {M\"uller}}, \bibinfo
  {author} {\bibfnamefont {E.}~\bibnamefont {Rico}}, \bibinfo {author}
  {\bibfnamefont {P.}~\bibnamefont {Stebler}}, \bibinfo {author} {\bibfnamefont
  {U.-J.}\ \bibnamefont {Wiese}}, \ and\ \bibinfo {author} {\bibfnamefont
  {P.}~\bibnamefont {Zoller}},\ }\href {\doibase
  10.1103/PhysRevLett.109.175302} {\bibfield  {journal} {\bibinfo  {journal}
  {Phys. Rev. Lett.}\ }\textbf {\bibinfo {volume} {109}},\ \bibinfo {pages}
  {175302} (\bibinfo {year} {2012})}\BibitemShut {NoStop}%
\bibitem [{\citenamefont {Banerjee}\ \emph {et~al.}(2013)\citenamefont
  {Banerjee}, \citenamefont {B\"ogli}, \citenamefont {Dalmonte}, \citenamefont
  {Rico}, \citenamefont {Stebler}, \citenamefont {Wiese},\ and\ \citenamefont
  {Zoller}}]{Banerjee2013}%
  \BibitemOpen
  \bibfield  {author} {\bibinfo {author} {\bibfnamefont {D.}~\bibnamefont
  {Banerjee}}, \bibinfo {author} {\bibfnamefont {M.}~\bibnamefont {B\"ogli}},
  \bibinfo {author} {\bibfnamefont {M.}~\bibnamefont {Dalmonte}}, \bibinfo
  {author} {\bibfnamefont {E.}~\bibnamefont {Rico}}, \bibinfo {author}
  {\bibfnamefont {P.}~\bibnamefont {Stebler}}, \bibinfo {author} {\bibfnamefont
  {U.-J.}\ \bibnamefont {Wiese}}, \ and\ \bibinfo {author} {\bibfnamefont
  {P.}~\bibnamefont {Zoller}},\ }\href {\doibase
  10.1103/PhysRevLett.110.125303} {\bibfield  {journal} {\bibinfo  {journal}
  {Phys. Rev. Lett.}\ }\textbf {\bibinfo {volume} {110}},\ \bibinfo {pages}
  {125303} (\bibinfo {year} {2013})}\BibitemShut {NoStop}%
\bibitem [{\citenamefont {Edmonds}\ \emph {et~al.}(2013)\citenamefont
  {Edmonds}, \citenamefont {Valiente}, \citenamefont
  {Juzeli\ifmmode~\bar{u}\else \={u}\fi{}nas}, \citenamefont {Santos},\ and\
  \citenamefont {\"Ohberg}}]{Edmonds2013}%
  \BibitemOpen
  \bibfield  {author} {\bibinfo {author} {\bibfnamefont {M.~J.}\ \bibnamefont
  {Edmonds}}, \bibinfo {author} {\bibfnamefont {M.}~\bibnamefont {Valiente}},
  \bibinfo {author} {\bibfnamefont {G.}~\bibnamefont
  {Juzeli\ifmmode~\bar{u}\else \={u}\fi{}nas}}, \bibinfo {author}
  {\bibfnamefont {L.}~\bibnamefont {Santos}}, \ and\ \bibinfo {author}
  {\bibfnamefont {P.}~\bibnamefont {\"Ohberg}},\ }\href {\doibase
  10.1103/PhysRevLett.110.085301} {\bibfield  {journal} {\bibinfo  {journal}
  {Phys. Rev. Lett.}\ }\textbf {\bibinfo {volume} {110}},\ \bibinfo {pages}
  {085301} (\bibinfo {year} {2013})}\BibitemShut {NoStop}%
\bibitem [{\citenamefont {Zohar}\ \emph
  {et~al.}(2013{\natexlab{a}})\citenamefont {Zohar}, \citenamefont {Cirac},\
  and\ \citenamefont {Reznik}}]{Zohar2013a}%
  \BibitemOpen
  \bibfield  {author} {\bibinfo {author} {\bibfnamefont {E.}~\bibnamefont
  {Zohar}}, \bibinfo {author} {\bibfnamefont {J.~I.}\ \bibnamefont {Cirac}}, \
  and\ \bibinfo {author} {\bibfnamefont {B.}~\bibnamefont {Reznik}},\ }\href
  {\doibase 10.1103/PhysRevA.88.023617} {\bibfield  {journal} {\bibinfo
  {journal} {Phys. Rev. A}\ }\textbf {\bibinfo {volume} {88}},\ \bibinfo
  {pages} {023617} (\bibinfo {year} {2013}{\natexlab{a}})}\BibitemShut
  {NoStop}%
\bibitem [{\citenamefont {Zohar}\ \emph
  {et~al.}(2013{\natexlab{b}})\citenamefont {Zohar}, \citenamefont {Cirac},\
  and\ \citenamefont {Reznik}}]{Zohar2013b}%
  \BibitemOpen
  \bibfield  {author} {\bibinfo {author} {\bibfnamefont {E.}~\bibnamefont
  {Zohar}}, \bibinfo {author} {\bibfnamefont {J.~I.}\ \bibnamefont {Cirac}}, \
  and\ \bibinfo {author} {\bibfnamefont {B.}~\bibnamefont {Reznik}},\ }\href
  {\doibase 10.1103/PhysRevLett.110.055302} {\bibfield  {journal} {\bibinfo
  {journal} {Phys. Rev. Lett.}\ }\textbf {\bibinfo {volume} {110}},\ \bibinfo
  {pages} {055302} (\bibinfo {year} {2013}{\natexlab{b}})}\BibitemShut
  {NoStop}%
\bibitem [{\citenamefont {Kasamatsu}\ \emph {et~al.}(2013)\citenamefont
  {Kasamatsu}, \citenamefont {Ichinose},\ and\ \citenamefont
  {Matsui}}]{Kasamatsu2013}%
  \BibitemOpen
  \bibfield  {author} {\bibinfo {author} {\bibfnamefont {K.}~\bibnamefont
  {Kasamatsu}}, \bibinfo {author} {\bibfnamefont {I.}~\bibnamefont {Ichinose}},
  \ and\ \bibinfo {author} {\bibfnamefont {T.}~\bibnamefont {Matsui}},\ }\href
  {\doibase 10.1103/PhysRevLett.111.115303} {\bibfield  {journal} {\bibinfo
  {journal} {Phys. Rev. Lett.}\ }\textbf {\bibinfo {volume} {111}},\ \bibinfo
  {pages} {115303} (\bibinfo {year} {2013})}\BibitemShut {NoStop}%
\bibitem [{\citenamefont {{Zohar}}\ \emph {et~al.}(2016)\citenamefont
  {{Zohar}}, \citenamefont {{Cirac}},\ and\ \citenamefont
  {{Reznik}}}]{Zohar2016}%
  \BibitemOpen
  \bibfield  {author} {\bibinfo {author} {\bibfnamefont {E.}~\bibnamefont
  {{Zohar}}}, \bibinfo {author} {\bibfnamefont {J.~I.}\ \bibnamefont
  {{Cirac}}}, \ and\ \bibinfo {author} {\bibfnamefont {B.}~\bibnamefont
  {{Reznik}}},\ }\href {\doibase 10.1088/0034-4885/79/1/014401} {\bibfield
  {journal} {\bibinfo  {journal} {Rep. Prog. Phys.}\ }\textbf {\bibinfo
  {volume} {79}},\ \bibinfo {eid} {014401} (\bibinfo {year}
  {2016})}\BibitemShut {NoStop}%
\bibitem [{\citenamefont {Hauke}\ \emph {et~al.}(2013)\citenamefont {Hauke},
  \citenamefont {Marcos}, \citenamefont {Dalmonte},\ and\ \citenamefont
  {Zoller}}]{Hauke2013}%
  \BibitemOpen
  \bibfield  {author} {\bibinfo {author} {\bibfnamefont {P.}~\bibnamefont
  {Hauke}}, \bibinfo {author} {\bibfnamefont {D.}~\bibnamefont {Marcos}},
  \bibinfo {author} {\bibfnamefont {M.}~\bibnamefont {Dalmonte}}, \ and\
  \bibinfo {author} {\bibfnamefont {P.}~\bibnamefont {Zoller}},\ }\href
  {\doibase 10.1103/PhysRevX.3.041018} {\bibfield  {journal} {\bibinfo
  {journal} {Phys. Rev. X}\ }\textbf {\bibinfo {volume} {3}},\ \bibinfo {pages}
  {041018} (\bibinfo {year} {2013})}\BibitemShut {NoStop}%
\bibitem [{\citenamefont {Tagliacozzo}\ \emph {et~al.}(2013)\citenamefont
  {Tagliacozzo}, \citenamefont {Celi}, \citenamefont {Zamora},\ and\
  \citenamefont {Lewenstein}}]{Tagliacozzo2013}%
  \BibitemOpen
  \bibfield  {author} {\bibinfo {author} {\bibfnamefont {L.}~\bibnamefont
  {Tagliacozzo}}, \bibinfo {author} {\bibfnamefont {A.}~\bibnamefont {Celi}},
  \bibinfo {author} {\bibfnamefont {A.}~\bibnamefont {Zamora}}, \ and\ \bibinfo
  {author} {\bibfnamefont {M.}~\bibnamefont {Lewenstein}},\ }\href {\doibase
  http://dx.doi.org/10.1016/j.aop.2012.11.009} {\bibfield  {journal} {\bibinfo
  {journal} {Ann. Phys. (N. Y.)}\ }\textbf {\bibinfo {volume} {330}},\ \bibinfo
  {pages} {160 } (\bibinfo {year} {2013})}\BibitemShut {NoStop}%
\bibitem [{\citenamefont {{Yang}}\ \emph {et~al.}(2016)\citenamefont {{Yang}},
  \citenamefont {{Shankar Giri}}, \citenamefont {{Johanning}}, \citenamefont
  {{Wunderlich}}, \citenamefont {{Zoller}},\ and\ \citenamefont
  {{Hauke}}}]{Yang2016}%
  \BibitemOpen
  \bibfield  {author} {\bibinfo {author} {\bibfnamefont {D.}~\bibnamefont
  {{Yang}}}, \bibinfo {author} {\bibfnamefont {G.}~\bibnamefont {{Shankar
  Giri}}}, \bibinfo {author} {\bibfnamefont {M.}~\bibnamefont {{Johanning}}},
  \bibinfo {author} {\bibfnamefont {C.}~\bibnamefont {{Wunderlich}}}, \bibinfo
  {author} {\bibfnamefont {P.}~\bibnamefont {{Zoller}}}, \ and\ \bibinfo
  {author} {\bibfnamefont {P.}~\bibnamefont {{Hauke}}},\ }\href@noop {}
  {\enquote {\bibinfo {title} {{Analog Quantum Simulation of (1+1)D Lattice QED
  with Trapped Ions}},}\ } (\bibinfo {year} {2016}),\ \Eprint
  {http://arxiv.org/abs/1604.03124} {arXiv:1604.03124} \BibitemShut {NoStop}%
\bibitem [{\citenamefont {Martinez}\ \emph {et~al.}(2016)\citenamefont
  {Martinez}, \citenamefont {Muschik}, \citenamefont {Schindler}, \citenamefont
  {Nigg}, \citenamefont {Erhard}, \citenamefont {Heyl}, \citenamefont {Hauke},
  \citenamefont {Dalmonte}, \citenamefont {Monz}, \citenamefont {Zoller},\ and\
  \citenamefont {Blatt}}]{martinez16}%
  \BibitemOpen
  \bibfield  {author} {\bibinfo {author} {\bibfnamefont {E.~A.}\ \bibnamefont
  {Martinez}}, \bibinfo {author} {\bibfnamefont {C.~A.}\ \bibnamefont
  {Muschik}}, \bibinfo {author} {\bibfnamefont {P.}~\bibnamefont {Schindler}},
  \bibinfo {author} {\bibfnamefont {D.}~\bibnamefont {Nigg}}, \bibinfo {author}
  {\bibfnamefont {A.}~\bibnamefont {Erhard}}, \bibinfo {author} {\bibfnamefont
  {M.}~\bibnamefont {Heyl}}, \bibinfo {author} {\bibfnamefont {P.}~\bibnamefont
  {Hauke}}, \bibinfo {author} {\bibfnamefont {M.}~\bibnamefont {Dalmonte}},
  \bibinfo {author} {\bibfnamefont {T.}~\bibnamefont {Monz}}, \bibinfo {author}
  {\bibfnamefont {P.}~\bibnamefont {Zoller}}, \ and\ \bibinfo {author}
  {\bibfnamefont {R.}~\bibnamefont {Blatt}},\ }\href {\doibase
  10.1038/nature18318} {\bibfield  {journal} {\bibinfo  {journal} {Nature
  (London)}\ }\textbf {\bibinfo {volume} {534}},\ \bibinfo {pages} {516}
  (\bibinfo {year} {2016})}\BibitemShut {NoStop}%
\bibitem [{\citenamefont {Marcos}\ \emph {et~al.}(2013)\citenamefont {Marcos},
  \citenamefont {Rabl}, \citenamefont {Rico},\ and\ \citenamefont
  {Zoller}}]{Marcos2013}%
  \BibitemOpen
  \bibfield  {author} {\bibinfo {author} {\bibfnamefont {D.}~\bibnamefont
  {Marcos}}, \bibinfo {author} {\bibfnamefont {P.}~\bibnamefont {Rabl}},
  \bibinfo {author} {\bibfnamefont {E.}~\bibnamefont {Rico}}, \ and\ \bibinfo
  {author} {\bibfnamefont {P.}~\bibnamefont {Zoller}},\ }\href {\doibase
  10.1103/PhysRevLett.111.110504} {\bibfield  {journal} {\bibinfo  {journal}
  {Phys. Rev. Lett.}\ }\textbf {\bibinfo {volume} {111}},\ \bibinfo {pages}
  {110504} (\bibinfo {year} {2013})}\BibitemShut {NoStop}%
\bibitem [{\citenamefont {Marcos}\ \emph {et~al.}(2014)\citenamefont {Marcos},
  \citenamefont {Widmer}, \citenamefont {Rico}, \citenamefont {Hafezi},
  \citenamefont {Rabl}, \citenamefont {Wiese},\ and\ \citenamefont
  {Zoller}}]{Marcos2014}%
  \BibitemOpen
  \bibfield  {author} {\bibinfo {author} {\bibfnamefont {D.}~\bibnamefont
  {Marcos}}, \bibinfo {author} {\bibfnamefont {P.}~\bibnamefont {Widmer}},
  \bibinfo {author} {\bibfnamefont {E.}~\bibnamefont {Rico}}, \bibinfo {author}
  {\bibfnamefont {M.}~\bibnamefont {Hafezi}}, \bibinfo {author} {\bibfnamefont
  {P.}~\bibnamefont {Rabl}}, \bibinfo {author} {\bibfnamefont {U.-J.}\
  \bibnamefont {Wiese}}, \ and\ \bibinfo {author} {\bibfnamefont
  {P.}~\bibnamefont {Zoller}},\ }\href {\doibase
  http://dx.doi.org/10.1016/j.aop.2014.09.011} {\bibfield  {journal} {\bibinfo
  {journal} {Ann. Phys. (N. Y.)}\ }\textbf {\bibinfo {volume} {351}},\ \bibinfo
  {pages} {634 } (\bibinfo {year} {2014})}\BibitemShut {NoStop}%
\bibitem [{\citenamefont {Papageorge}\ \emph {et~al.}(2016)\citenamefont
  {Papageorge}, \citenamefont {Koll\'{a}r},\ and\ \citenamefont
  {Lev}}]{Papageorge2016}%
  \BibitemOpen
  \bibfield  {author} {\bibinfo {author} {\bibfnamefont {A.~T.}\ \bibnamefont
  {Papageorge}}, \bibinfo {author} {\bibfnamefont {A.~J.}\ \bibnamefont
  {Koll\'{a}r}}, \ and\ \bibinfo {author} {\bibfnamefont {B.~L.}\ \bibnamefont
  {Lev}},\ }\href {\doibase 10.1364/OE.24.011447} {\bibfield  {journal}
  {\bibinfo  {journal} {Opt. Express}\ }\textbf {\bibinfo {volume} {24}},\
  \bibinfo {pages} {11447} (\bibinfo {year} {2016})}\BibitemShut {NoStop}%
\bibitem [{\citenamefont {Schmidt}\ \emph {et~al.}(2016)\citenamefont
  {Schmidt}, \citenamefont {Mayer}, \citenamefont {Hohmann}, \citenamefont
  {Lausch}, \citenamefont {Kindermann},\ and\ \citenamefont
  {Widera}}]{Schmidt:2016hv}%
  \BibitemOpen
  \bibfield  {author} {\bibinfo {author} {\bibfnamefont {F.}~\bibnamefont
  {Schmidt}}, \bibinfo {author} {\bibfnamefont {D.}~\bibnamefont {Mayer}},
  \bibinfo {author} {\bibfnamefont {M.}~\bibnamefont {Hohmann}}, \bibinfo
  {author} {\bibfnamefont {T.}~\bibnamefont {Lausch}}, \bibinfo {author}
  {\bibfnamefont {F.}~\bibnamefont {Kindermann}}, \ and\ \bibinfo {author}
  {\bibfnamefont {A.}~\bibnamefont {Widera}},\ }\href@noop {} {\bibfield
  {journal} {\bibinfo  {journal} {Phys. Rev. A}\ }\textbf {\bibinfo {volume}
  {93}},\ \bibinfo {pages} {022507} (\bibinfo {year} {2016})}\BibitemShut
  {NoStop}%
\bibitem [{\citenamefont {LeBlanc}\ \emph {et~al.}(2015)\citenamefont
  {LeBlanc}, \citenamefont {Beeler}, \citenamefont {Jim{\'e}nez-Garc{\'\i}a},
  \citenamefont {Williams}, \citenamefont {Phillips},\ and\ \citenamefont
  {Spielman}}]{LeBlanc2015}%
  \BibitemOpen
  \bibfield  {author} {\bibinfo {author} {\bibfnamefont {L.~J.}\ \bibnamefont
  {LeBlanc}}, \bibinfo {author} {\bibfnamefont {M.~C.}\ \bibnamefont {Beeler}},
  \bibinfo {author} {\bibfnamefont {K.}~\bibnamefont
  {Jim{\'e}nez-Garc{\'\i}a}}, \bibinfo {author} {\bibfnamefont {R.~A.}\
  \bibnamefont {Williams}}, \bibinfo {author} {\bibfnamefont {W.~D.}\
  \bibnamefont {Phillips}}, \ and\ \bibinfo {author} {\bibfnamefont {I.~B.}\
  \bibnamefont {Spielman}},\ }\href@noop {} {\bibfield  {journal} {\bibinfo
  {journal} {New J. Phys.}\ }\textbf {\bibinfo {volume} {17}},\ \bibinfo
  {pages} {065016} (\bibinfo {year} {2015})}\BibitemShut {NoStop}%
\bibitem [{Note1()}]{Note1}%
  \BibitemOpen
  \bibinfo {note} {Supplemental material, containing details of reduction to
  effective 2D equations, details of numerical simulation, and discussion of
  artificial magnetic field strength suppression in the geometry under
  consideration.}\BibitemShut {Stop}%
\bibitem [{\citenamefont {Siegman}(1986)}]{Siegman1986a}%
  \BibitemOpen
  \bibfield  {author} {\bibinfo {author} {\bibfnamefont {A.~E.}\ \bibnamefont
  {Siegman}},\ }\href@noop {} {\emph {\bibinfo {title} {{Lasers}}}}\ (\bibinfo
  {publisher} {University Science Books},\ \bibinfo {address} {Sausolito},\
  \bibinfo {year} {1986})\BibitemShut {NoStop}%
\bibitem [{\citenamefont {Dennis}\ \emph {et~al.}(2013)\citenamefont {Dennis},
  \citenamefont {Hope},\ and\ \citenamefont {Johnsson}}]{Dennis2013201}%
  \BibitemOpen
  \bibfield  {author} {\bibinfo {author} {\bibfnamefont {G.~R.}\ \bibnamefont
  {Dennis}}, \bibinfo {author} {\bibfnamefont {J.~J.}\ \bibnamefont {Hope}}, \
  and\ \bibinfo {author} {\bibfnamefont {M.~T.}\ \bibnamefont {Johnsson}},\
  }\href {\doibase http://dx.doi.org/10.1016/j.cpc.2012.08.016} {\bibfield
  {journal} {\bibinfo  {journal} {Comput. Phys. Commun.}\ }\textbf {\bibinfo
  {volume} {184}},\ \bibinfo {pages} {201 } (\bibinfo {year}
  {2013})}\BibitemShut {NoStop}%
\end{thebibliography}
%

\appendix
\renewcommand{\theequation}{S\arabic{equation}}
\setcounter{equation}{0}
\renewcommand{\thefigure}{S\arabic{figure}}
\setcounter{figure}{0}

\clearpage

\onecolumngrid
\begin{center}
\textbf{\large Supplemental Materials}
\end{center}
\vspace{\columnsep}
\twocolumngrid

In this supplemental material, we discuss the reduction from the full three-dimensional problem of atoms in an anisotropic trap coupled to cavity modes,
to the effective two-dimensional equations for atoms and cavities.  We also
present further details of the numerical simulations, and discuss further the nature of
the magnetic field suppression in the two-dimensional geometry we consider.

\section{Reduction to effective two-dimensional equations}

To derive the equations of motion Eqs.~(\ref{phieom}) and~(\ref{psieom}),
we need to use the structure of the cavity modes to eliminate the $z$
dependence. 
The intensity of light in the 
cavity can be expressed as a sum of cavity eigenmodes $u_\mu$, 
\begin{equation}
\hat{I} = \left|\sum \hat{a}_\mu u_\mu(\mathbf{r},z) \right|^2,
\end{equation}
where the mode functions take the form
\begin{equation}
u_\mu(\ver,z)=\chi_\mu(\ver) \cos(g(z)+\theta(z) (\mu_x+\mu_y)).
\end{equation}
Here $\chi_\mu$ are normalized Hermite-Gauss modes with index $\mu_{x(y)}$ in
the $x(y)$ directions respectively, i.e., 
$\chi_\mu=\chi_{\mu_x}^x(\sqrt{2}x/w)\chi_{\mu_y}^y(\sqrt{2}y/w)/w$ with
\begin{equation}
\chi_{\mu_x}^x=\sqrt[4]{\frac{2}{\pi}}\frac{1}{\sqrt{2^{\mu_x}\mu_x!}}H_{\mu_x}(x)e^{-x^2},
\end{equation}
and similarly for $y$, where $w=w_0\sqrt{1+(z/z_R)^2}$ is the beam waist and $H_n$ is the
$n\mathrm{th}$ Hermite function.
$g(z)=k[z+r^2/2R(z)]-\Phi(z)$ 
in terms of the Gouy phase $\Phi(z) = \arctan(z/z_R)$~\cite{Siegman1986a},
the radius of curvature is $R(z)=z+z_R^2/z$, and
$\theta(z)=\Phi(z)+\pi/4$.
Here $z_R$ is the Rayleigh range, $z_R = k w_0^2/2$.
The mode amplitudes $\alpha_\mu=\langle\hat{a}_\mu\rangle$ 
follow the equation of motion
$i\partial_t \alpha_\mu=\left\langle\left[\hat{a}_\mu,\hat{H}\right]
\right\rangle-i\kappa\alpha_\mu$, where $\hat{H}=\hat{H}_\text{atom}+\hat{H}_\text{cav}+f_\mu(\hat{a}_\mu^\dagger+\hat{a}_\mu)$
and $f_\mu$ is the overlap of the pump beam with each mode. 
This gives
\begin{equation}
\label{modeeom}
\begin{aligned}
i\partial_t \alpha_\mu &=(\omega_\mu-\omega_P-i\kappa)\alpha_\mu+f_\mu
- P_{\mu\nu}\alpha_\nu \\
P_{\mu\nu} &=\frac{1}{2}\int \dif^2\vec{r} \dif z\, u_\mu u_\nu(\e_A\abs{\Psi_A}^2+\e_B\abs{\Psi_B}^2),
\end{aligned}
\end{equation}
where $\omega_\mu$ is the frequency of mode $\mu$.

As discussed in the main text, we consider a pancake of atoms by writing 
$\Psi_{A,B}(\ver,z)=\psi_{A,B}(\ver) Z(z)$ where
\begin{equation}
\abs{Z(z)}^2=\frac{1}{\sqrt{2\pi\sigma_z^2}}
\exp\left(-\frac{(z-z_0)^2}{2\sigma_z^2}\right).
\end{equation}
We can then eliminate the $z$ dependence from all the above
terms, and write effective transverse dynamics of atoms and light.
To proceed, we  compute the integral
\begin{equation}
I_{\perp}(\vec{r}) =\int \dif z\, \abs{Z(z)}^2 I(\ver,z).
\end{equation}
To do this, we work in the approximation that
$\lambda\ll\sigma_z\ll z_R$ so that we can drop the fast oscillating terms
depending on $g(z)$ and evaluate $\theta(z)$ at $z=z_0$. This gives
\begin{multline}
\label{intz}
\int \dif z\, \abs{Z(z)}^2 u_\mu(\ver,z) u_\nu(\ver,z) \approx \\
 \frac{1}{2}\chi_\mu(\ver)\chi_\nu(\ver)
\cos(\theta(z_0)(\mu_x+\mu_y-\nu_x-\nu_y)).
\end{multline}
This allows us to write the dynamics of $\psi_{A,B}(\vec r)$ in terms of
the cavity mode amplitudes $\alpha_\mu$.

Turning to the equations for the cavity mode amplitudes, we first
define the transverse field as
\begin{equation}
\varphi(\ver)=\sum \alpha_\mu \chi_\mu(\ver).
\end{equation}
With this definition, we
multiply Eq.~(\ref{modeeom}) by $\chi_\mu(\ver^\prime)$ and sum over $\mu$ to
get the equation of motion for $\varphi(\ver^\prime)$.  In order to be able to
write the transverse dynamics of the light field in real space, we  use
the fact that Hermite-Gauss modes are eigenmodes of the harmonic
oscillator Hamiltonian
\begin{equation}
\omega_\mu \chi_\mu(\ver)=
\left[\frac{\delta}{2}\left(
    -l^2\nabla^2+\frac{r^2}{l^2}\right)+\omega_0\right]\chi_\mu(\ver),
\end{equation} 
where the beam waist is $\sqrt{2}l$, $\omega_0$
is the frequency of the confocal cavity, and $\delta$ is the mode spacing
when the cavity is not perfectly confocal.

To close the equations, we must write $I_{\perp}(\vec r)$ in terms
of $\varphi(\vec r)$.
For a general $z_0$, the light in the
cavity will mediate non-local interactions between the atoms, and
so this involves a convolution. However,
we will assume that the atoms are close to the end of the
cavity, which in the confocal case is at $z_0=-z_R$. This leads to
a quasi-local matter-light interaction, and thus, as we see in the
main text, provides a close analog to the standard Meissner effect. 
Since $\theta(-z_R)=0$, the cosine term
drops out of Eq.~(\ref{intz}), and
so $I_{\perp}(\vec r)=|\varphi(\vec r)|^2/2$.
Together with the relation
\begin{equation}
\sum \chi_\mu(\ver) \chi_\mu(\ver^\prime) = \delta(\ver-\ver^\prime),
\end{equation}
this leads to the equation of motion for the 2D
cavity field, which combined with the equation of motion of the atom
condensate, describes the dynamics of the system.

\section{Further details of numerical simulation}

In order to have an approximately constant artificial magnetic field 
over a large area while keeping the
intensity symmetric, we choose a pump profile
\begin{equation}
f(\ver)=f_0\sqrt{\abs{y}}\sqrt{\kappa^2+(\delta r^2/2l^2-\Delta_0)^2}
\end{equation}
which, in an empty cavity, would give $|\varphi|^2 \approx \abs{f_0}^2\abs{y}$.
We then convolve this profile with a Gaussian of width $0.5$ so that it is
smoothed near $y=0$. Hence $B_z \approx \pm\abs{f_0}^2$, as shown in
Fig.~\ref{fig2}(c). Such a
profile can be achieved using, e.g., a digital multimirror
device (DMD)~\cite{Papageorge2016}.

\section{Nature of suppression of magnetic field inside atom cloud}

As noted in the main text, there are some distinctions between the behaviour we predict here, and the standard Meissner effect.  This section discusses in further detail the origin of these distinctions.

In the standard Meissner effect, there is a vector potential $A_\mu$ which couples (via minimal coupling)
to the matter component.  The remaining action for the system then takes the Maxwell form, $F_{\mu\nu}F^{\mu\nu}$, in terms of the field tensor
$F_{\mu\nu} = \partial_\mu A_\nu - \partial_\nu A_\mu$.
Our equations of motion correspond to an action that differs in several ways.  Firstly, our action is non-relativistic (i.e., not Lorenz covariant): this affects the dynamics, but not the steady state.  Secondly, our action is not gauge invariant---i.e., our action depends directly on $|\varphi|^2$, which plays the role of the vector potential, rather than depending only on derivatives of this quantity.  Most notably, our action is not gapless.  This last distinction, which occurs due to the inevitable cavity loss terms,  means that there is no sharp distinction between the case of coupling to superfluid or normal atoms:  the magnetic field always has a finite penetration depth---i.e., away from the location of the source term $f(\vec{r})$, the field $\varphi$ will decay.  However, in our numerical results, we clearly see a dramatic change in the suppression of the magnetic field.  This is consistent, as the effect we are seeing can be understood as arising from a dramatic enhancement of the effective gap.  Explicitly, in Eq.~\ref{phieom}, the superfluid induced gap, $N\e_{\Delta}(\abs{\psi_A}^2-\abs{\psi_B}^2)$, is at least an order of magnitude larger than the other terms
$\kappa$ and $\Delta_0$, leading to a large suppression of the magnetic field.

An additional difference from textbook Meissner physics arises because of the distinct geometry of our system.  In our geometry, both the atom cloud, and the synthetic field vary only in the transverse two-dimensional plane.  In the textbook example of the Meissner example, a piece of superconducting material is placed in an externally imposed uniform magnetic field.  However, imposing an uniform field requires sources, e.g., loops of free current placed far away in the $z$ direction.  In our geometry, with the synthetic field restricted to two dimensions, this is not possible.  The source of the vector potential (in our case the longitudinal pumping term, $f(\vec{r})$), must live in the same plane as the material.  Generating a pseudo uniform magnetic field thus requires an extended source term.  The specific case
we consider has this source exist across the entire plane.  Thus, we cannot expect complete expulsion of the magnetic field, but only suppression.

\begin{figure}[t!]
  \centering
   \includegraphics[width=\columnwidth]{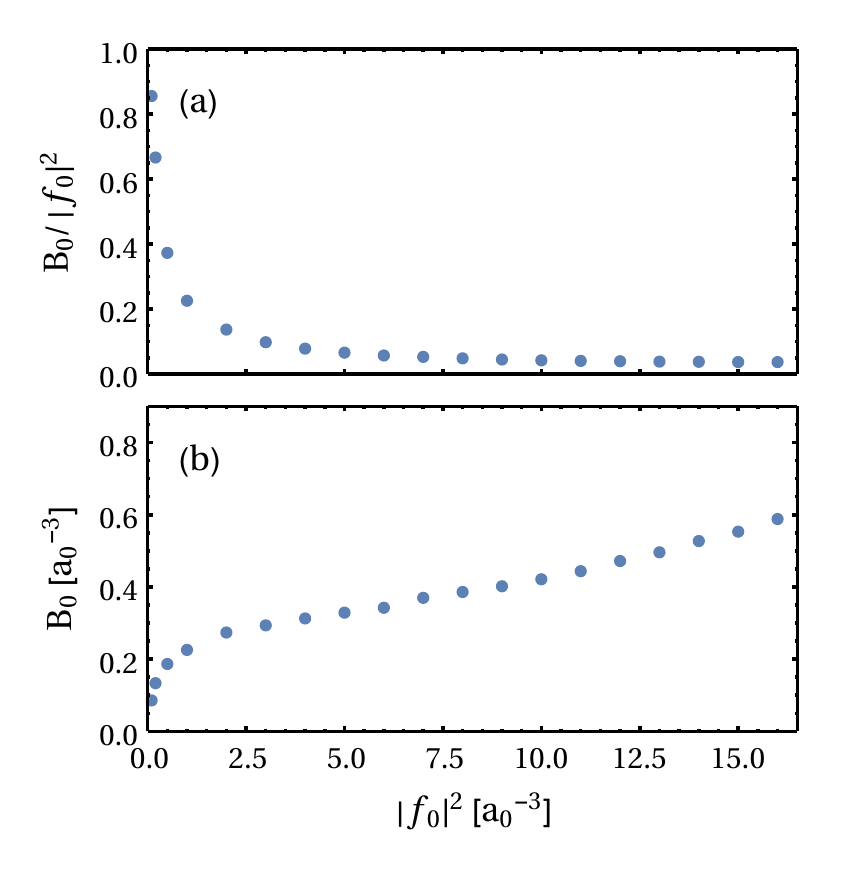}
  \caption{Plateau value of magnetic field inside condensate (a) normalized
  by pump intensity and (b) unnormalized as a function of pump intensity. 
  Increasing pump strength leads to a stronger diamagnetic response so that
  the relative field falls and the absolute value of the field increases sub
  linearly for small pump intensities. We choose $NU=7500$ and all other
  parameters are the same as Fig.~\ref{fig2}.
  }
  \label{figs1}
\end{figure}

To explore the above issues further numerically,
Fig.~\ref{figs1} shows the artificial magnetic field inside the
condensate (at $x=0$, $y/a_0=1$) as a function of pump intensity (or
equivalently, applied field strength). 
Most terms in Eq.~\ref{phieom} are linear in $\varphi$. However the diamagnetic
response, according to Eq.~\ref{feedback}, is proportional to $\abs{\varphi}^2$. 
Hence, for very low pump strength it is approximately zero and the resulting
field is equal to the applied field. As the pump intensity increases, the
diamagnetic term begins to dominate. One thus observes that the relative field strength falls by several orders of magnitude. Equivalently, the absolute field strength increases
sub-linearly.

\end{document}